\documentclass[11pt,a4paper]{article}
\usepackage[a4paper,margin=1in]{geometry}
\usepackage{graphicx}
\usepackage{amsmath}
\usepackage{amssymb}
\usepackage{authblk}
\usepackage{hyperref}
\usepackage{caption}
\usepackage[section]{placeins}
\usepackage{booktabs}
\usepackage{subcaption}
\usepackage{indentfirst}
\usepackage{lineno}
\usepackage{xcolor}
\usepackage{makecell}
\usepackage[comma,numbers,square,sort&compress]{natbib} 

\usepackage{comment}

\title{\bfseries Comparison of Image Processing Models in Quark--Gluon Jet Classification}
\author{Daeun Kim}
\author{Jaeyoon Cho}
\author{Jiwon Lee}
\author{Wonjun Jeong}
\author{Hyeongwoo Noh}
\author{Giyeong Kim}
\author{Seunghwan Yang}
\author{MinJung Kweon\thanks{Corresponding author: \texttt{minjung@inha.ac.kr}}}
\affil{Department of Physics, Inha University, Incheon, Republic of Korea}
\date{}

\begin{document}
\maketitle

\begin{abstract}
\noindent

Quark–gluon discrimination provides a useful test case for studying how different machine-learning architectures learn the spatial structure of QCD radiation. In this work, we compare convolutional neural network (CNN), Vision Transformers (ViT), and hierarchical Swin Transformers using the same three-channel jet-image representation, consisting of charged-particle momentum, neutral-particle momentum, and charged-particle multiplicity from PYTHIA 8 jets. We study their performance for different training-set sizes and fine-tuning configurations, with particular attention to the role of local and global information in the jet images.
CNN and Swin models consistently perform better than ViT in the cases studied. Since both CNN and Swin retain a strong local component in their architectures, this suggests that local jet substructure plays an important role in quark–gluon discrimination. The performance of the hierarchical Swin model also suggests that combining local features over larger spatial scales is useful. Block-wise fine-tuning improves the performance of the Transformer models, although the improvement becomes smaller and the training less stable as more blocks are unfrozen. We also find that self-supervised Momentum Contrast (MoCo) pretraining improves the model initialization, particularly when the amount of labeled training data is limited.
Based on these observations, we developed a smaller Swin model adopted to the jet-image representation used in this study. It achieves comparable performance with substantially fewer parameters.
The results show that it is important to adapt the model architecture and training procedure to the specific input characteristics of High Energy Physics (HEP) data when applying vision models in HEP.

\end{abstract}

\section{Introduction}

Jets are collimated sprays of hadrons resulting from the fragmentation and hadronization of high-energy quarks or gluons produced in high-energy collisions.
At the Large Hadron Collider (LHC), the study of jet substructure provides crucial insights into parton shower dynamics, color coherence, and quark--gluon plasma properties. Since the early 2010s, machine learning has been increasingly used to classify jets based on internal energy flow patterns rather than physics-motivated observables~\cite{Larkoski2017}.

The classification of quark and gluon jets is particularly challenging due to their similar overall shapes and stochastic parton shower processes.
However, the difference in color factors --- $C_F=4/3$ for quarks and $C_A=3$ for gluons --- leads to different particle multiplicities and radiation patterns:
\begin{equation}
    \frac{\langle n_{\mathrm{tr}} \rangle_g}
    {\langle n_{\mathrm{tr}} \rangle_q}
    \approx \frac{C_A}{C_F} \approx 2.25,
\end{equation}
where $\langle n_{\mathrm{tr}} \rangle$ denotes the average number of charged particle tracks~\cite{JHEP2017}.
These differences are reflected in the spatial and momentum distributions of jet constituents and can therefore be represented in jet images.

In this work, we study quark--gluon jet classification using Convolutional Neural Network (CNN) and transformer-based vision models, specifically the Vision Transformer (ViT) and Shifted Window (Swin) Transformer.
CNN extracts local features through convolutional kernels~\cite{cnn1989}, whereas ViT uses self-attention between image patches~\cite{vit2020}.
The Swin Transformer restricts attention to local windows and introduces a hierarchical structure through shifted windows and patch merging, allowing information to be combined over progressively larger spatial scales~\cite{swin2021}.
This provides an interesting comparison for jet images, where discriminating information can be present both in localized substructure and in radiation patterns over larger angular scales.

Modern jet tagging has increasingly moved toward models that operate directly on particle constituents.
Particle-cloud and graph-based models, Lorentz-equivariant architectures, and constituent-based Transformers such as ParT have achieved state-of-the-art performance without projecting the constituent information onto a regular image grid~\cite{particleCloudTransformer,Esser2021,efnPFN,He2023PDAT,lorentzNet2022,particleTransformer}.
Our aim is to study how different vision architectures and training choices behave under the same conditions using a common jet-image representation instead of establishing the superiority of image-based models over modern constituent-based approaches.
Image-based methods are also relevant to  High Energy Physics (HEP) data that naturally have a regular spatial structure, such as calorimeter or LArTPC data, where an image-like representation follows more directly from the detector readout.

Because pretrained vision models are generally developed using natural-image datasets, their application to jet images also raises the question of how much of the pretrained model should be fine-tuned.
We therefore study block-wise fine-tuning for ViT and Swin and examine its dependence on the available training statistics.
We also investigate self-supervised pretraining using Momentum Contrast (MoCo)~\cite{moco2020}.
These approaches provide ways to learn useful representations from unlabeled data before the final classification task.

Using a common three-channel jet-image representation, we compare CNN, ViT, and Swin models over several training-set sizes and fine-tuning configurations.
We also develop a smaller Swin model that operates directly on the original $72\times72$ jet-image resolution and apply MoCo pretraining to this model.
The goal of the custom model is to study whether adapting the model size and input resolution to the jet-image data can provide comparable performance with reduced computational requirements, rather than introducing a new state-of-the-art jet tagger.

The present study focuses on a controlled comparison of different architectures and training strategies using truth-level simulated jets.
Detector response, pile-up, generator dependence, and simulation-to-data differences are not included.
These effects can be important for experimental applications, particularly for image-based models, and will need to be studied with detector-level simulation and ultimately real data.
A direct comparison with modern constituent-based models under the same training and computational conditions is also left for future work.

This paper is organized as follows.
Section 2 reviews related machine-learning approaches to jet tagging.
Section 3 describes the quark--gluon jet image dataset used in this analysis.
Section 4 introduces the employed machine-learning models, including CNN, ViT, Swin Transformer, and the MoCo framework.
Section 5 details the training setup, fine-tuning strategies, and customized Swin architecture.
Section 6 presents the classification performance and its dependence on sample size and fine-tuning depth.
Finally, Section 6 summarizes our conclusions and discusses future prospects.

\section{Related works}
Machine-learning approaches to jet tagging differ in how jet information is represented and processed.
Below, we briefly review representative methods relevant to quark--gluon classification, organized according to their input representations and model architectures.

\begin{itemize}
    \item \textbf{Image-Based Models.}

    Image-based models use visual representations constructed from jet data as input. Representative approaches to quark--gluon classification include Jet Images and their CNN-based extensions~\cite{Cogan2015,Oliveira2016}, the multi-channel representation introduced in \textit{Deep Learning in Color}~\cite{Komiske2017}, and Vision Transformer-based classification using calorimeter images~\cite{Jahin2025ViTJets}.

    \item \textbf{Set-Based Constituent Models.}

    Set-based constituent models treat a jet as a variable-length, unordered set of its constituent particles. Representative examples include Energy Flow Networks (EFNs) and Particle Flow Networks (PFNs)~\cite{efnPFN}, which use permutation-invariant architectures to perform jet classification from constituent-level information.

    \item \textbf{Graph-Based Constituent Models.}

    Graph-based constituent models represent a jet as a graph of its constituent particles and learn relationships among the particles through graph operations. ParticleNet applies dynamic graph convolutions to particle-cloud representations~\cite{particleNet}, whereas LorentzNet uses Minkowski dot-product attention to aggregate information among constituents while preserving Lorentz equivariance~\cite{lorentzNet2022}.

    \item \textbf{Constituent-Based Transformer Models.}

    Constituent-based Transformer models use attention mechanisms to process low-level features of jet constituents and learn correlations among them. Representative architectures include the Point Cloud Transformer (PCT)~\cite{particleCloudTransformer}, Particle Transformer (ParT)~\cite{particleTransformer}, Particle Dual Attention Transformer (P-DAT)~\cite{He2023PDAT}, and Lorentz-equivariant L-GATr~\cite{spinner2024lorentz}. Transformer-based taggers developed for experimental applications include CMS UParT~\cite{UParTCMS} and ATLAS GN2~\cite{atlas2026transforming} for jet-flavour tagging, as well as ATLAS DeParT for quark--gluon tagging~\cite{DeParTAtlas}.

\end{itemize}

It should also be noted that the application of image-based models in HEP is not limited to jet classification. CNN-based approaches have also been studied for trigger applications, including FPGA-based implementations for Level-0 triggers and event classification for the High-Luminosity Large Hadron Collider (HL-LHC), where small network architectures are particularly important due to strict latency and resource~\cite{fpgaHEP, Brooke:2025cdc}.

Within this broader landscape, the present study focuses on image-based representations and examines how convolutional and Transformer-based vision architectures, including models pretrained on natural and object-centric images, can be adapted to quark--gluon classification. Using a common jet-image representation, we compare CNN, ViT, and Swin architectures, with particular attention to fine-tuning depth, training dataset size, and self-supervised pretraining. This scope enables a controlled methodological comparison of architectural and training choices within the image-based setting.

\section{Dataset}

We use the open dataset \textit{Pythia8 Quark and Gluon Jets for Energy Flow}, in which jets are generated with \textsc{Pythia~8} at a center-of-mass energy of $\sqrt{s}=14~\mathrm{TeV}$ and reconstructed using the anti-$k_{\rm T}$ algorithm with $R=0.4$~\cite{zenodo}.
In this analysis, the dataset is restricted to light-flavor quark jets, excluding charm and bottom quark jets.
The selected jets have transverse momentum $p_{\rm T}^{\mathrm{jet}} \in [500,550]~\mathrm{GeV}$ and rapidity $|y_{\mathrm{jet}}|<1.7$.

Each jet is represented as a $72\times72$ pixel image in $(y,\phi)$ space, where $y$ and $\phi$ denote the rapidity and azimuthal angle, respectively.
We follow the colored jet-image procedure proposed in~\cite{Komiske2017}, including jet centering and grid discretization.
The jet images use three channels:
\begin{itemize}
    \item \textbf{R-channel:} $p_{\rm T}$ of charged particles;
    \item \textbf{G-channel:} $p_{\rm T}$ of neutral particles;
    \item \textbf{B-channel:} charged-particle multiplicity.
\end{itemize}
The R and G channels are normalized by the total jet $p_{\rm T}$, obtained by summing the R and G pixel values over the full image, while the B channel is left unnormalized.
This representation retains both momentum and charged-particle multiplicity information in a form that can be used directly as input to image-based models.

The present study uses truth-level particles without detector simulation or pile-up. The $72\times72$ binning is used as a fixed representation for the comparison of the different models and is not intended to reproduce the granularity of a specific detector.
Detector effects such as finite spatial and momentum resolution, reconstruction efficiency, and detector-induced fluctuations can modify the resulting image and are not considered in the present study.
Their impact on the model performance will be investigated in future work using detector-level simulations.

\begin{figure}[htbp]
\centering
\begin{subfigure}{0.45\linewidth}
\centering
\includegraphics[height=4cm,keepaspectratio]{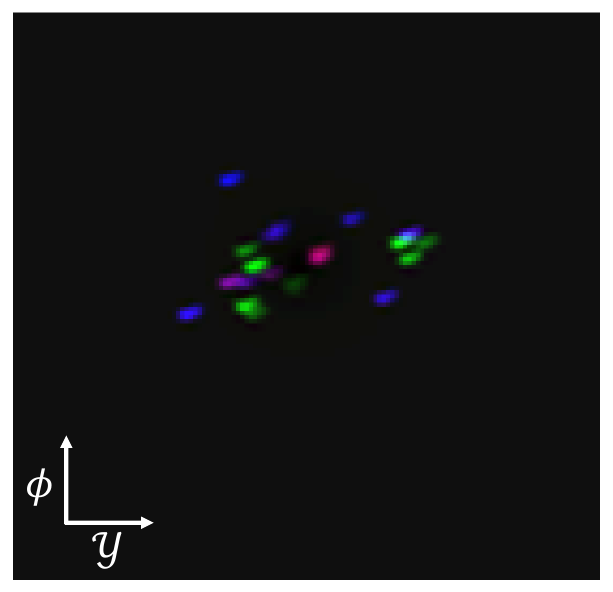}
\caption{Quark jet}
\label{fig:quark}
\end{subfigure}
\hfill
\begin{subfigure}{0.45\linewidth}
\centering
\includegraphics[height=4cm,keepaspectratio]{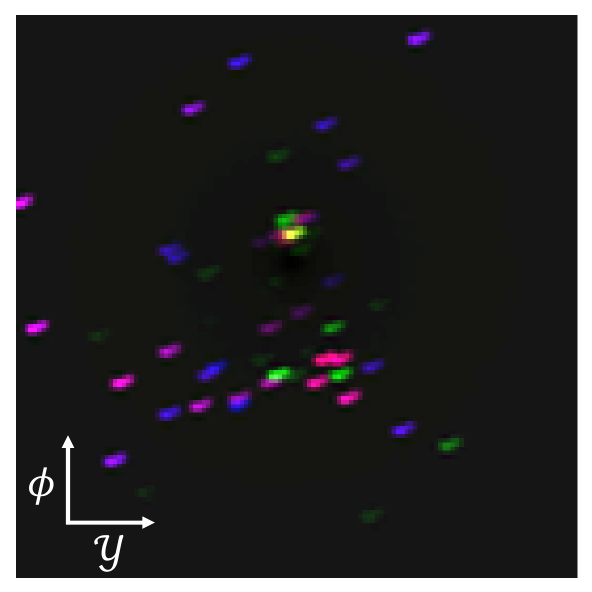}
\caption{Gluon jet}
\label{fig:gluon}
\end{subfigure}
\caption{
Examples of (a) a quark jet and (b) a gluon jet represented as three-channel jet images.
For visualization purposes, the colored pixels are displayed larger than their actual size in the $72\times72$ images.
}
\label{fig:quark_gluon}
\end{figure}

Examples of the resulting three-channel jet images for quark and gluon jets are shown in Figure~\ref{fig:quark_gluon}.

\section{Model Architectures}

\subsection{Convolutional Neural Network}

CNN extracts features by sliding fixed-size convolutional filters across an image
and computing features from the pixel patches at each location~\cite{cnn1989}.
Since the output of each neuron depends only on the pixels within the filter's receptive region, CNN is structurally most sensitive to correlations between nearby pixels~\cite{Oliveira2016}.
This characteristic can be viewed as a locality bias, meaning that CNN features are constructed primarily from local neighborhoods by design.
Due to this structural property, CNN effectively captures local spatial patterns in jet images by emphasizing correlations among nearby pixels, which aligns well with fine-grained jet substructure patterns~\cite{Komiske2017, Cogan2015}.
In this sense, CNN is particularly effective at modeling short-range spatial dependencies.

However, the region observable by a single convolutional layer is limited by the filter size~\cite{DumoulinVisinConvArithmetic}.
To learn relationships between distant regions---such as between the jet core and outer regions---the network must pass information through multiple convolutional layers, progressively expanding the receptive field of higher layers~\cite{Hesse2023CAD}. Moreover, the effective receptive field typically occupies only a fraction of the theoretical receptive field, with pixel contributions concentrated near the center in an approximately Gaussian manner~\cite{Luo2016}.
This process introduces a design trade-off: if the network depth is insufficient or spatial resolution decreases rapidly due to downsampling at intermediate stages, fine details can be attenuated, making it harder to maintain informative long-range correlations across the entire jet image~\cite{Hesse2023CAD}.

CNN is strong at capturing short-range patterns, but they have difficulty explicitly modeling long-range interactions~\cite{NonLocal2018}.
Because jet image classification can involve both local substructure and more global organization of radiation patterns, it is useful to consider architectures that can balance local inductive bias with global context modeling.
This locality bias and the way receptive fields expand in CNN motivates a comparison with Transformer-based models that employ self-attention mechanisms~\cite{Esser2021, Raghu2021}.

\subsection{Vision Transformer}

ViT represents an image as a sequence of tokens and feeds it into a Transformer encoder~\cite{vit2020}.
The input image is split into fixed-size patches, and each patch is linearly projected into a patch embedding.
For classification tasks, a class token is prepended to the sequence to obtain a representative global representation, and positional embeddings are added to each token to encode spatial location information within the image.
The Transformer encoder then applies multiple layers of self-attention blocks to model interactions among tokens~\cite{Transformer2017}.

Let the embedding matrix consisting of $N$ patch tokens be $X \in \mathbb{R}^{N \times d}$, where $d$ is the embedding dimension.
For self-attention computation, each token is linearly projected into query ($Q$), key ($K$), and value ($V$) as follows~\cite{Transformer2017}:
\begin{equation}
Q = XW_Q, \quad K = XW_K, \quad V = XW_V, \quad X \in \mathbb{R}^{N \times d},
\end{equation}
where $W_Q, W_K, W_V$ are learnable weight matrices that project the input embeddings into query, key, and value vectors.

Transformer self-attention computes token-to-token similarity via the dot product between query and key, normalizes it using softmax function to obtain attention weights, and updates each token representation by taking a weighted sum of the value vectors. To mitigate the issue that similarity scores grow with dimensionality, ViT adopts the standard  scaled dot-product attention, where $d_k$ denotes the dimensionality of the projected query/key vectors:
\begin{equation}
\mathrm{Attention}(Q,K,V)=\mathrm{softmax}\!\left(\frac{QK^{\top}}{\sqrt{d_k}}\right)V.
\end{equation}

\begin{figure}[htbp]
    \centering
    \includegraphics[width=0.8\linewidth]{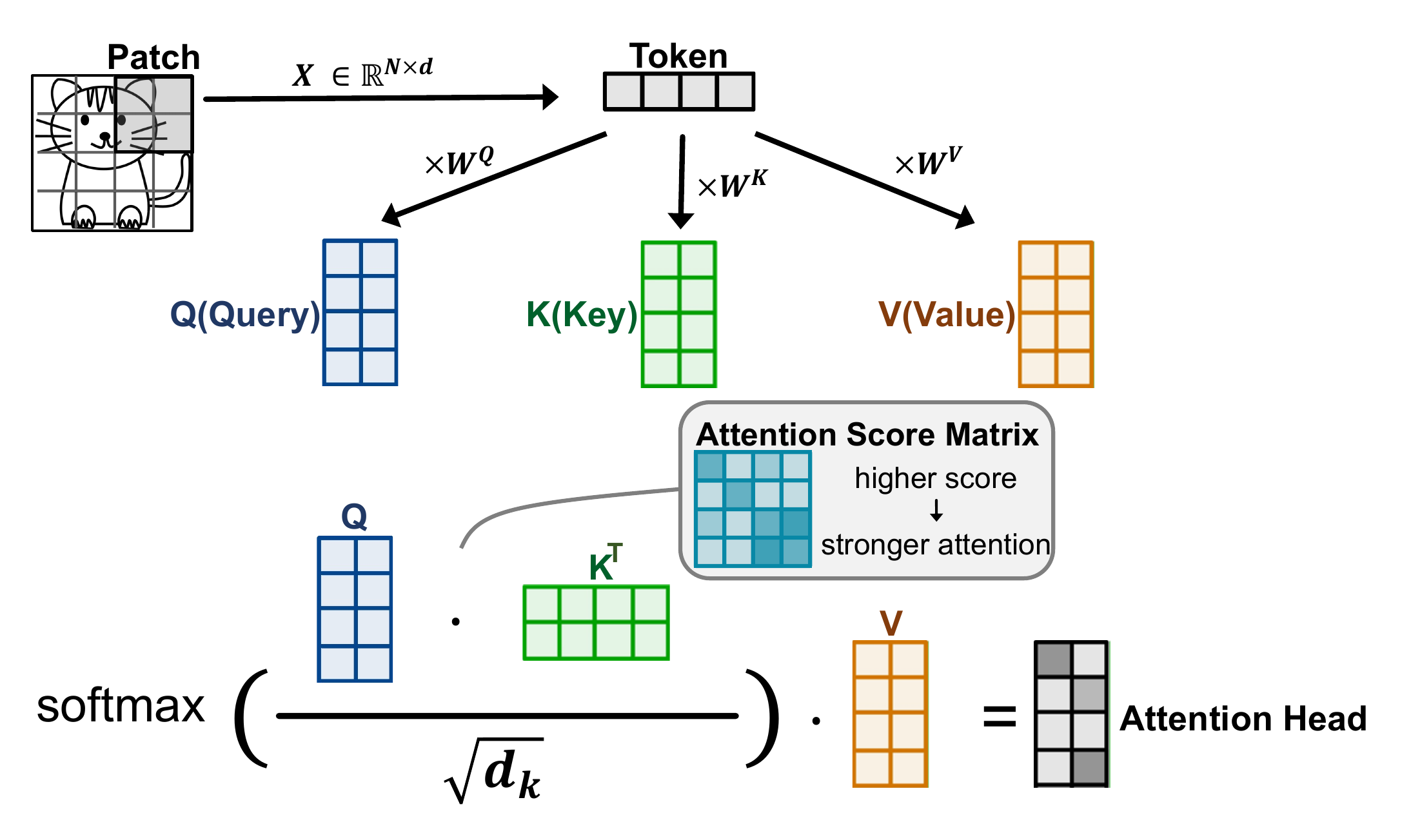}
    \caption{Simplified schematic of the self-attention mechanism. A $4 \times 4 $ image is split into $2 \times 2$ patches, which are projected into query, key and value embeddings. Attention scores are computed and applied to obtain the output representation for a single attention head. The class token is omitted for clarity. The attention score matrix represents pairwise query-key similarities, where higher scores indicate stronger attention.}
    \label{fig:attention}
\end{figure}

A single attention operation estimates relationships within one projection subspace, which can limit expressiveness.
To improve representational capacity, ViT employs multi-head attention, generating multiple sets of $Q$, $K$, and $V$ with different learned weights~\cite{Transformer2017}. In this setting, the projected query/key dimensionality $d_k$ in Eq. (3) corresponds to the per-head dimension $d_{head}$. Attention is computed in parallel across heads and the results are concatenated:
\begin{equation}
\mathrm{MultiHead}(X) = [\mathrm{head}_1,\ldots,\mathrm{head}_h]W_O.
\end{equation}
Figure~\ref{fig:attention} is a schematic illustration of the computational flow of the multi-head self-attention defined in Eqs. (3)-(4). Since each head follows the same computational procedure with different learned weights, the figure presents a representative instance of the per-head attention operation.

Standard ViT uses global self-attention, which can directly model relationships between all pairs of tokens.
As a result, it can integrate global information across the entire image from the early stages of learning.
These structural properties can be advantageous for jet image classification~\cite{Jahin2025ViTJets}. In jet images, classification-relevant information is not confined to local pixel-level patterns; it may also involve long-range dependencies such as the relationship between energy distributions in the jet core and outer regions, the global organization of radiation patterns, and correlation structures spanning large radii. This suggests that such long-range dependencies can be directly captured via token-to-token attention.

In addition, ViT uses positional embeddings so that token order and spatial location are not lost during attention computation.
Given that jet images are constructed on an $(y,\phi)$ grid, positional embeddings help the model preserve where each token originates in the $(y,\phi)$ space~\cite{vit2020}, enabling it to learn global interactions while still respecting spatial structure.
Consequently, ViT can achieve global context-centric representation learning by combining information across multiple spatial scales---such as overall jet radius and radiation patterns---through attention-based aggregation.

\subsection{Swin Transformer}

CNN-based approaches have been applied to jet image classification in prior studies~\cite{Oliveira2016, Komiske2017}.
While CNN effectively captures patterns in neighboring regions through convolution operations, ViT can directly integrate global information across an image via self-attention~\cite{vit2020}.
However, a standard ViT may exhibit a relatively weaker inductive bias for learning local patterns~\cite{CoAtNet2021}.
Therefore, in this study, we adopt the Swin Transformer as a practical compromise that preserves the benefit of modeling global interactions while strengthening feature extraction in local regions~\cite{swin2021, Yang2021Focal}.

The key idea of the Swin Transformer is \emph{window-based attention}, where self-attention is computed locally within non-overlapping windows rather than globally over the entire patch sequence.
This operation is referred to as window-based multi-head self-attention (W-MSA). For an $M \times M$ window, the attention operation defined in Eq. (3) is applied independently to the $M^2$ tokens within each window, so that each token interacts only with other tokens in the same local window.

This design enables efficient learning of local features.
In the subsequent block, the windows are shifted to allow cross-window communication.
This operation is referred to as shifted-window multi-head self-attention (SW-MSA). For an $M \times M$ window, the window partition is shifted by $\lfloor M/2 \rfloor$ patches along each spatial dimension relative to the preceding W-MSA block. In practice, this shift is implemented using a cyclic shift. Since the cyclic shift can place spatially non-adjacent regions within the same attention window through wrap-around, an attention mask is applied to prevent such artificial interactions. The same window-based attention is then computed within the shifted windows after masking, allowing tokens that belonged to different neighboring windows during W-MSA to directly interact during SW-MSA. W-MSA and SW-MSA are alternated across successive blocks, enabling information to propagate across neighboring windows while retaining local attention computation.

In addition, the Swin Transformer repeatedly applies a \emph{patch merging} layer between stages.
Let $H$ and $W$ denote the number of patch tokens along the height and width directions, respectively, and let $C$ denote the dimension of each token representation. The feature vectors of each group of neighboring $2 \times 2$ tokens are concatenated, reducing the number of tokens along each spatial dimension by a factor of two while increasing the token dimension from $C$ to $4C$. A linear projection is then applied to reduce the concatenated $4C$-dimensional representation to $2C$. The patch merging operation can therefore be expressed as
\begin{equation}
H \times W \times C \rightarrow \frac{H}{2} \times \frac{W}{2} \times 4C \rightarrow \frac{H}{2} \times \frac{W}{2} \times 2C
\end{equation}
Thus, patch merging reduces the number of spatial tokens by a factor of four while increasing the token dimension. This operation is applied between stages, whereas W-MSA and SW-MSA operate within each stage at a fixed spatial resolution. The merged representation then serves as the input to the attention blocks of the subsequent stage.
As this procedure is repeated through successive stages, the number of spatial tokens decreases while each token represents information from a progressively larger region of the original image.
The Swin Transformer therefore forms a hierarchical representation in which early stages retain relatively fine spatial information, while later stages describe features over larger spatial scales.
This differs from the standard ViT architecture, which maintains the same patch resolution throughout its Transformer blocks.
Such a hierarchical structure allows the model to combine local details with information over progressively larger regions~\cite{swin2021}.
Accordingly, the Swin Transformer can be considered a meaningful candidate model for jet classification, as it can simultaneously account for local pattern learning and broader context in jet images.

\begin{figure}[htbp]
    \centering
    \includegraphics[width=1.0\linewidth]{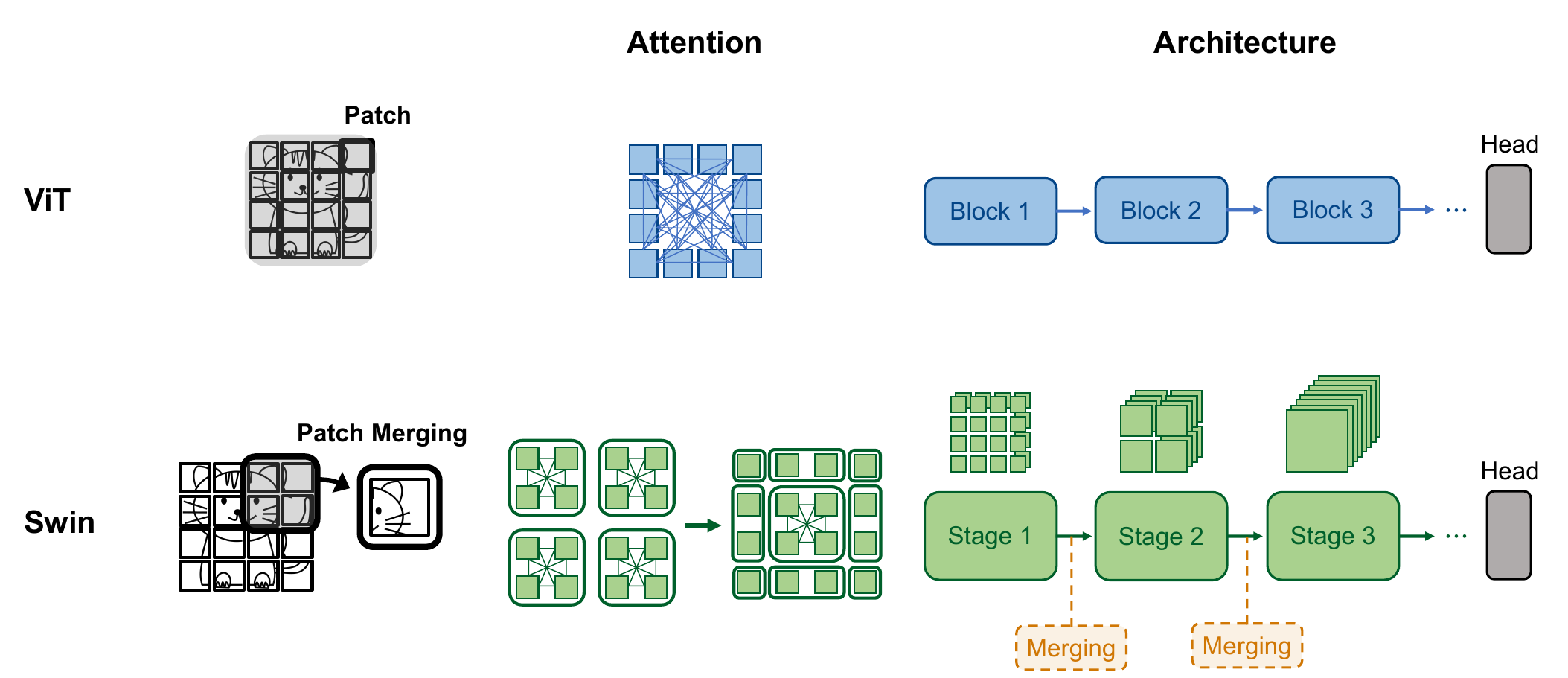}
    \caption{
    Schematic of structural differences between ViT and Swin Transformer. ViT employs global self-attention, whereas the Swin Transformer uses window-based local self-attention and shifted windows to enable cross-window interactions. Patch merging between successive stages halves the spatial resolution in each dimension and doubles the number of channel.
    }
    \label{fig:structural_difference}
\end{figure}

Figure~\ref{fig:structural_difference} summarizes the structural differences between ViT and Swin Transformer.
While ViT processes a fixed-resolution patch representation through a flat sequence of Transformer blocks, Swin organizes the blocks into successive stages connected by patch merging.
The resulting hierarchy gradually changes the representation from fine-grained local features in the early stages to broader spatial features in the later stages.

\subsection{Self-Supervised Momentum Contrast Pretraining}

Transformer-based architectures such as ViT and Swin typically require large-scale datasets to achieve strong performance when trained from scratch~\cite{vit2020,swin2021}. 
Consequently, these models are commonly initialized with weights pretrained on large natural-image datasets such as ImageNet to obtain robust feature representations~\cite{smallvit2021, imagenet21k2021}. 
However, jet images differ substantially from natural images in both structure and semantics, which introduces a domain discrepancy between the pretraining data and the target task~\cite{robusttransfer2020}. 

Self-supervised learning has been studied in recent jet-learning literature as an alternative way of obtaining pretrained representations. Examples include contrastive methods such as JetCLR and RS3L, masking-based methods such as Masked Particle Modeling (MPM), and predictive joint-embedding methods such as J-JEPA~\cite{symmetryssl2021,harris2025resimulation, golling2024masked, katel2024learning}. These approaches learn jet representations using different self-supervised signals, including data augmentation, re-simulation, masked-particle prediction, and latent representation prediction.

\begin{figure}[htbp]
    \centering
    \includegraphics[width=0.8\linewidth]{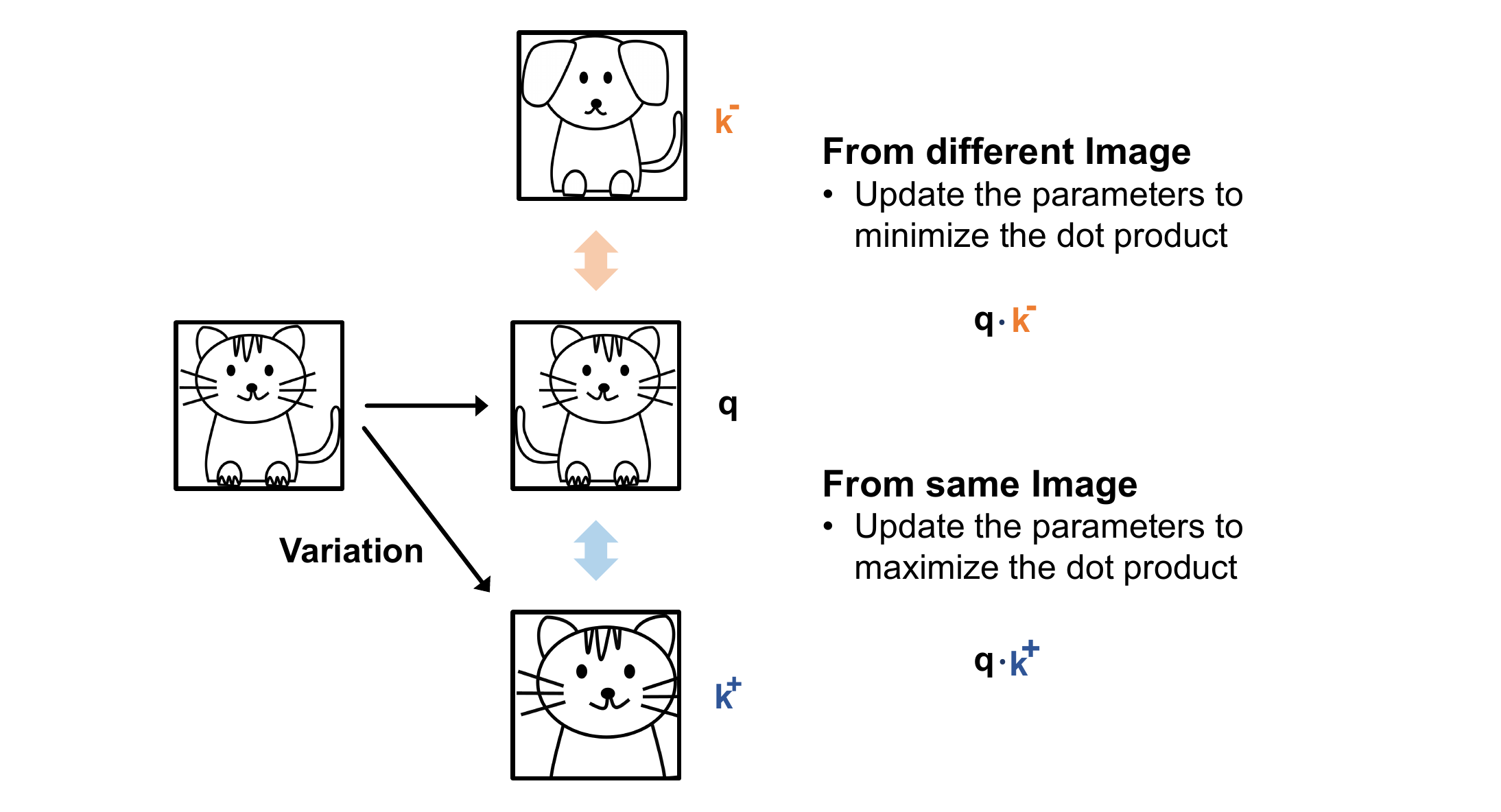}
    \caption{Schematic of the MoCo framework. An input image is transformed into two augmented views. One view is processed by the query encoder to produce the query $q$, while the other view is processed by the momentum encoder to produce the positive key $k^{+}$. Features from other images serve as negative keys $k^{-}$, which are stored in a dynamic queue. The model is trained to maximize $q \cdot k^{+}$ while minimizing $q \cdot k^{-}$.}
    \label{fig:MoCo_explain}
\end{figure}

In this study, we adopt MoCo~\cite{moco2020}, a self-supervised momentum contrastive learning framework that maintains a large and consistent set of negative samples using a momentum encoder. 
This design enables stable contrastive representation learning and makes MoCo a suitable choice for domain-specific pretraining on jet images.
The mechanism of MoCo is illustrated in Figure~\ref{fig:MoCo_explain}.

For each input image, two augmented views are generated through simple image transformations. One view is passed through the query encoder to produce a feature vector $q$, referred to as a query.
The other view is processed by the momentum encoder to produce a feature vector $k^{+}$, which acts as the positive key corresponding to the same image. 
In addition, features from other images are treated as negative keys, denoted as  $k^{-}$.
The model is trained to maximize the dot product $q \cdot k^{+}$ while minimizing $q \cdot k^{-}$.
During training, the query encoder that produces $q$ is updated by backpropagation using this contrastive objective. In contrast, the momentum encoder is not updated by backpropagation.
Instead, its parameters are updated as an exponential moving average of the encoder parameters. This update allows the momentum encoder to follow the query encoder with a slowly changing update, which stabilizes the representations used for contrastive learning.
After  this pretraining, the query encoder is used to initialize the model for the downstream task.~\cite{moco2020}

To generate different views for contrastive learning, we apply horizontal and vertical flips, rotations by multiples of $90^\circ$, and Gaussian blur. These transformations generate different views of the same jet image while preserving its overall structure. Detailed augmentation settings are provided in~\ref{app:moco}.

\section{Fine-Tuning and Custom Model Strategies}

\paragraph{Training setup}
All models were trained on an NVIDIA RTX 4060 GPU using CUDA on a Windows operating system.
Training was performed for 80 epochs with a batch size of 128, and models were selected based on the best validation ROC-AUC score.

For experiments using 20k and 200k jet images, the datasets were split into training and validation sets with a ratio of 8:2, while the 1.6M jet image dataset was split at a ratio of 9:1.
For evaluation, 10 independent test sets were constructed, each containing 20k jet images, with no overlap between the training, validation, and test sets.
No data augmentation or additional normalization was applied.
For transformer-based models pretrained on ImageNet, input images were upsampled from $72 \times 72$ to $224 \times 224$ during dataset loading.
Table~\ref{tab:training_setup} summarizes the model-specific training configurations.

\begin{table}[t]
\centering
\caption{Model-specific training configurations. Block-level tuning was applied to ViT and Swin. A large learning rate was used for ViT and Swin with 0 blocks since only the classification head is trained. The CNN also required a large rate for stable convergence.}
\label{tab:training_setup}
\begin{tabular}{lccccc}
\hline
\textbf{Model} & 
\textbf{Input size} &
\textbf{Learning rate} &
\textbf{Trainable blocks} &
\textbf{Params (M)}   \\
\hline
CNN &  $72 \times 72$ & $1\times10^{-3}$ & All layers & 1.42\\

ViT & $224 \times 224$ & $1\times10^{-5}$ (0 blocks: $1\times10^{-4}$) &
Last 0 / 2 / 4 & 5.52\\

Swin & $224 \times 224$ & $1\times10^{-5}$  (0 blocks: $1\times10^{-4}$) & Last 0 / 2 / 4 & 27.52\\

Custom Swin & $72 \times 72$ & $1\times10^{-5}$ & All blocks & 3.82\\

\hline
\end{tabular}
\end{table}

\subsection{Fine-Tuning Strategy}

Pretrained \texttt{vit\_tiny\_patch16\_224}(ViT-Tiny) and \texttt{swin\_tiny\_patch4\_window7\_224}(Swin-Tiny) models from the \texttt{timm} library~\cite{timm}, which are lightweight variants with reduced model size, originally trained on the ImageNet-1k dataset~\cite{vit2020,swin2021}, were fine-tuned on the jet image dataset.
Hereafter, the results obtained from fine-tuned ViT-Tiny and fine-tuned Swin-Tiny models are referred to as ViT and Swin, respectively.
Considering training cost and the risk of overfitting, a block-wise fine-tuning strategy was adopted~\cite{vit2020,swin2021,Yosinski2014}, in which only a subset of transformer blocks was unfrozen, allowing the parameters of the unfrozen blocks to be updated during training, while the remaining blocks were kept frozen and their parameters were not updated.

\begin{figure}[htbp]
    \centering
    \includegraphics[width=0.8\linewidth]{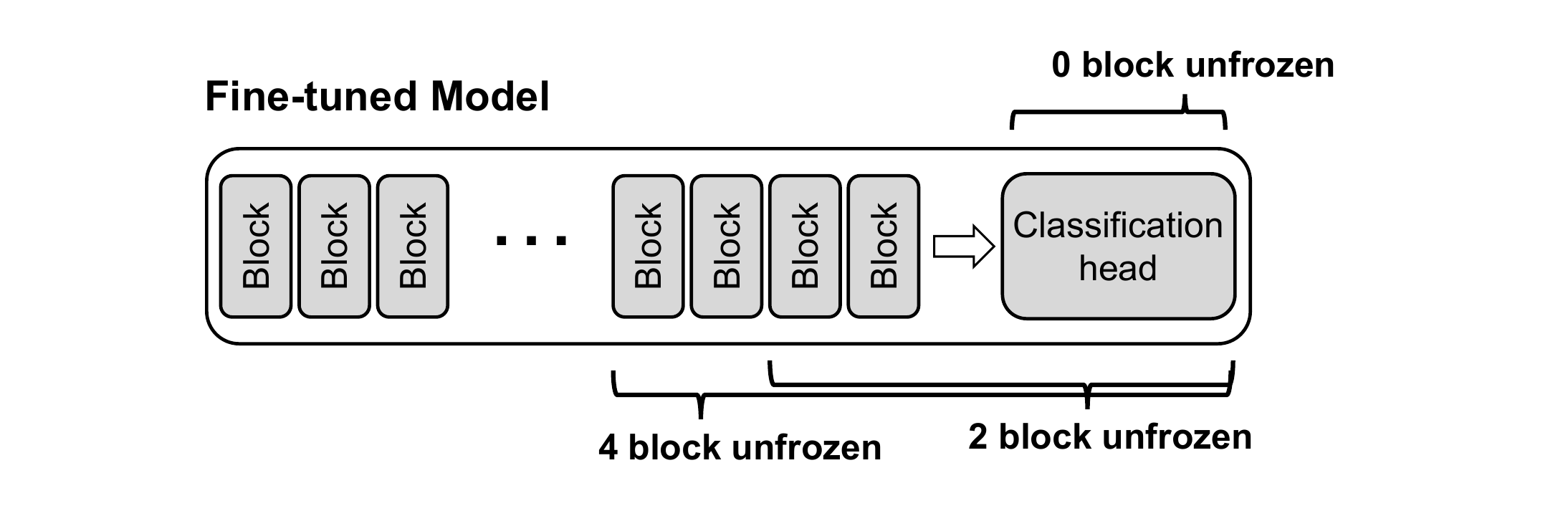}
    \caption{Schematic of the block-wise fine-tuning strategy applied to the pretrained ViT and Swin models.
    The number of unfrozen transformer blocks is progressively increased from the classification head, while the remaining blocks are kept frozen.}
    \label{fig:ft_explain}
\end{figure}

As illustrated in Figure~\ref{fig:ft_explain}, this strategy progressively unfreezes only the final transformer blocks closest to the classification head, while earlier blocks remain frozen, as deeper layers are known to encode more task-specific representations whereas earlier layers capture more general features that can be preserved during adaptation.
This block-wise fine-tuning scheme is applied identically to both ViT and Swin models, and Figure~\ref{fig:ft_explain} provides a schematic representation common to both architectures.

Since the task involves binary classification between quark jets and gluon jets, the classification head was left unfrozen in all configurations to adapt the ImageNet-pretrained 1,000-class output layer to a two-class setting, while the number of unfrozen transformer blocks was varied.
As shown in Table~\ref{tab:training_setup}, we explore three fine-tuning configurations for ViT and Swin, each differing in the number of blocks subjected to fine-tuning.

\begin{itemize}
    \item \textbf{0 blocks:} all transformer blocks frozen; only the classification head is trained;
    \item \textbf{2 blocks:} the last two transformer blocks unfrozen;
    \item \textbf{4 blocks:} the last four transformer blocks unfrozen.
\end{itemize}

\FloatBarrier

\subsection{Custom Swin Model Strategy}

Custom Swin model was implemented based on the Swin architecture,
with modifications to the input resolution and model configuration to better suit jet image data. Transformer-based models such as ViT and Swin are typically pretrained on natural images with a fixed input resolution of $224 \times 224$~\cite{vit2020,swin2021}.
However, jet images are originally defined on a lower-resolution $(y,\phi)$ grid~\cite{Esser2021}, so using such pretrained models requires upsampling, which introduces redundant pixels that do not contribute meaningful information for model training.

\begin{figure}[htbp]
    \centering
    \includegraphics[width=0.8\linewidth]{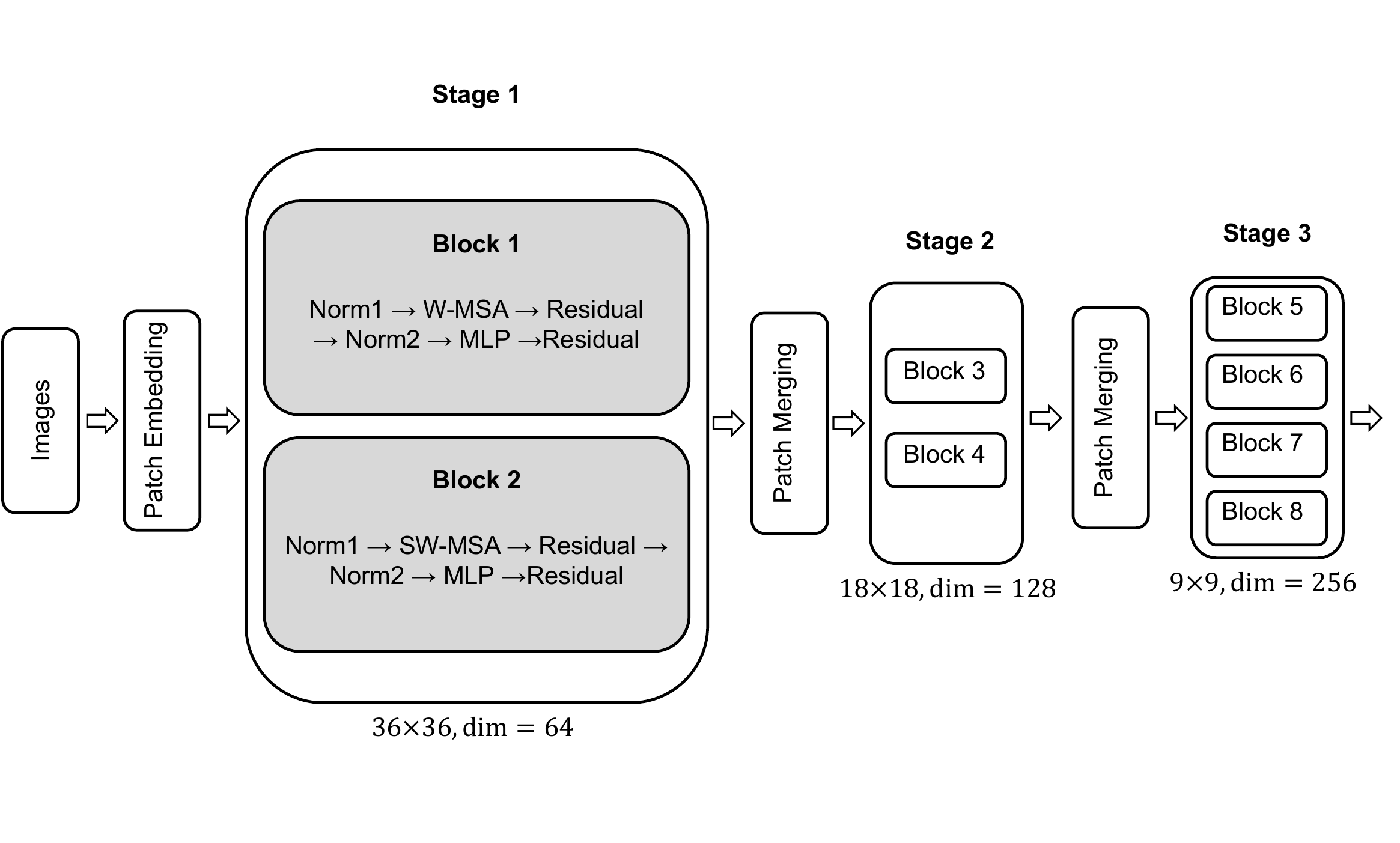}
    \caption{Schematic architecture of the custom Swin used in this study. Starting from the input image, tokens are generated through patch embedding. As the tokens pass through three hierarchical stages, the token resolution progressively decreases from $36\times36$ to $18\times18$ and $9\times9$, while the embedding dimension increases from 64 to 128 and 256. Norm denotes Layer Normalization, W-MSA denotes window-based multi-head self-attention, SW-MSA denotes shifted-window multi-head self-attention, and MLP denotes a multilayer perceptron. Each block follows the structure Norm $\rightarrow$ Attention $\rightarrow$ Residual $\rightarrow$ Norm $\rightarrow$ MLP $\rightarrow$ Residual. 
    The attention mechanism alternates between W-MSA and SW-MSA across consecutive blocks. Block A and Block B share the same structure across stages, with only token resolution and embedding dimension varying. Further implementation details of the model are provided in Appendix A.2.4.}
    \label{fig:custom_model}
\end{figure}

To address this issue, the input resolution was fixed to $72 \times 72$, matching the native resolution of the jet images.
This resolution remains sufficient to capture relevant jet substructure information while significantly reducing computational overhead.
In addition to the input resolution, the model capacity was adjusted to account for the limited computational resources and the binary nature of the classification task.
Transformer-based models with a large number of parameters are more prone to overfitting in low-data regimes~\cite{vit2020}.
To mitigate this issue, a custom Swin architecture was designed by reducing model parameters while preserving the hierarchical structure of the Swin (original Swin-Tiny) Transformer.
In addition to the input resolution, key architectural parameters were adapted to be consistent with the reduced input size, including a window size of $3 \times 3$, a patch size of $2 \times 2$, and a three-stage hierarchy with block depths of (2, 2, 4) and embedding dimension of (64, 128, 256).
As a result of these architectural modifications, the total number of model parameters was reduced by approximately 86\% compared to the original Swin architecture, leading to improved training efficiency. The overall architecture of the custom Swin model is illustrated in Figure~\ref{fig:custom_model}.

For self-supervised pretraining, the query encoder and momentum encoder in the MoCo framework (Figure~\ref{fig:MoCo_explain}) are implemented using the custom Swin architecture introduced above.
To obtain a jet-domain pretrained model comparable to the ViT and Swin models pretrained on general image datasets, the custom Swin architecture was further pretrained with MoCo using 200k jet image dataset used to train the CNN, ViT and Swin models.
After pretraining, the query encoder is used to initialize the backbone of the custom Swin model and is subsequently fine-tuned.

Training was conducted on datasets containing 20k and 200k samples using the Adam optimizer~\cite{kingma2014adam} while minimizing the cross-entropy loss \cite{Goodfellow2016}. 
The hyperparameters and training configurations used for each model are summarized in Table~\ref{tab:training_setup}. Detailed descriptions of the training procedure for each model are provided in the \ref{app:modelHypParam}.
To further investigate the impact of the training dataset size, several models were additionally trained using a larger dataset containing 1.6M samples.

\section{Results}

We computed the variation in model performance across multiple independent test sets as follows. 
For the 20k dataset, the data were split into training and validation sets with an 8:2 ratio (16k/4k), and the same ratio was used for the 200k dataset (160k/40k).
For the 1.6M dataset, the training and validation sets were split in a 9:1 ratio (1.44M/160k).
For evaluation, 10 independent test sets were constructed, each containing 20k jet images, with no overlap between the training, validation, and test sets.
The same test sets were used across all models to ensure a fair comparison.
Model performance was averaged over the 10 test sets, and the error bars represent the standard deviation ($1\sigma$).
Uncertainties related to the training procedure are not considered in this analysis.
The impact of the training seed is investigated separately for two selected models with the results discussed in a later section.

\subsection{Performance Comparison Across Models}

\begin{figure}[htb]
  \centering
  \includegraphics[width=\linewidth]{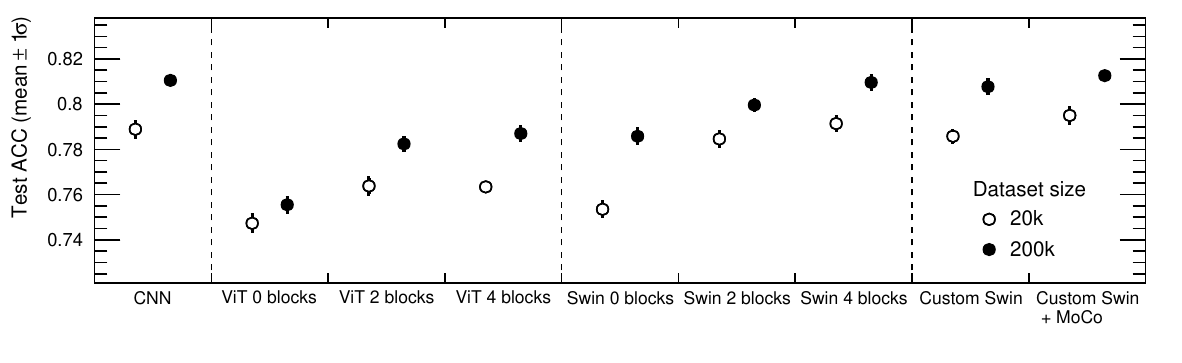}
  \includegraphics[width=\linewidth]{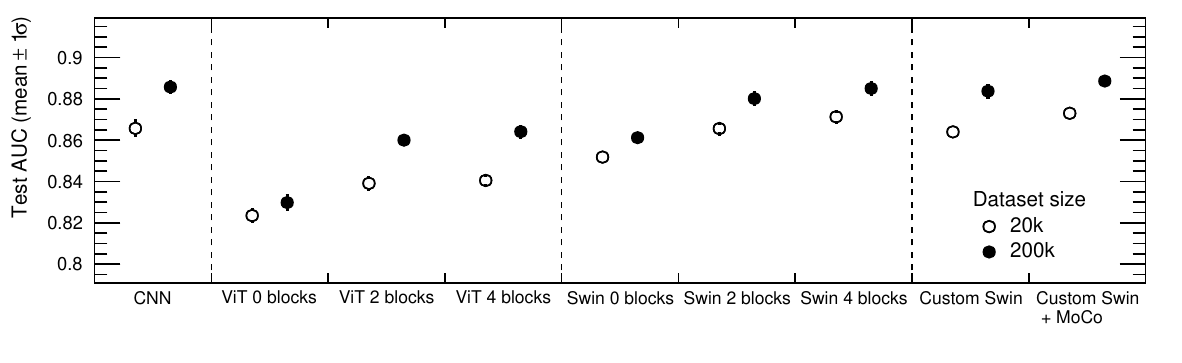}
  \caption{Test ACC (top) and AUC (bottom) for different model architectures trained on 20k and 200k jet image datasets~\cite{Komiske2017}, with models separated by dashed lines.
  Markers indicate mean performance and error bars denote one standard deviation obtained from the test set variation.
  CNN and Swin models show better performance compared to ViT models, while custom Swin models achieve comparable results with fewer parameters.}
\label{fig:all_test}
\end{figure}

Figure~\ref{fig:all_test} compares the classification performance of different model architectures in terms of accuracy (ACC) and the area under the ROC curve (AUC).
Increasing the dataset size from 20k to 200k consistently improves performance across all architectures. This suggests that model performance is not yet fully saturated and may benefit from larger training statistics, as a larger dataset provides more diverse training examples for model training.
The CNN and fine-tuned Swin models show better performance than ViT models for both 20k and 200k datasets.
While ViT are capable of learning localized patterns through self-attention, they lack an explicit inductive bias toward locality. In contrast, CNN and Swin architectures enforce local feature extraction in the early stages, which can be advantageous for jet images where discriminative information is often encoded in fine-grained substructure.
This suggests that an architectural bias toward local and hierarchical feature extraction may be beneficial for jet-image classification compared with relying primarily on global token interactions.
The custom Swin models, introduced to address the limitations of fine-tuned models, closely follow the performance trends of fine-tuned Swin architectures in both ACC and AUC, achieving performance comparable to that of the fine-tuned models despite using substantially fewer parameters (Table~\ref{tab:training_setup}).
These results show that the observed comparable performance comes from overcoming practical limitations of fine-tuned models, such as unnecessary image resizing and additional computation required to match fixed input sizes.
This suggests that, rather than simply reusing pretrained models as they are, designing models that are better suited to the jet image data can serve as an effective alternative for enabling the model to operate efficiently.

\begin{figure}[hbt]
  \centering
  \includegraphics[width=7.7cm]{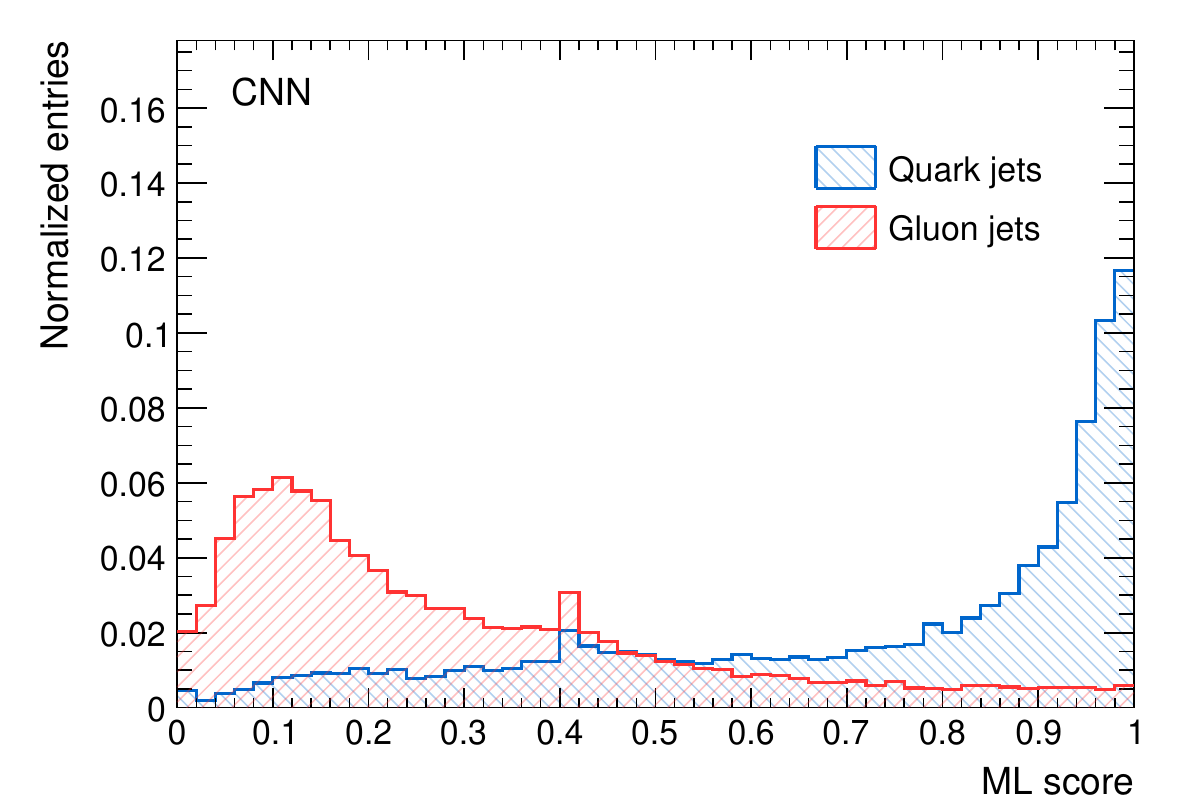}
  \includegraphics[width=7.7cm]{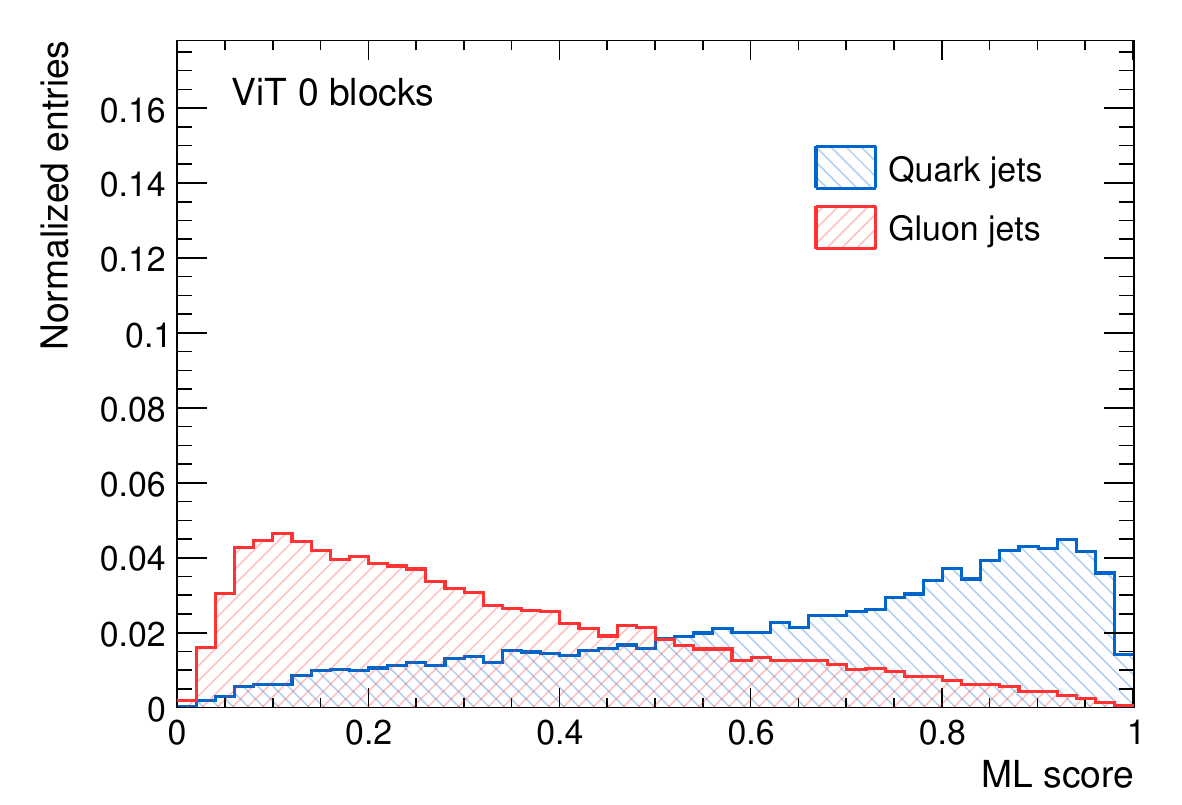}
  \includegraphics[width=7.7cm]{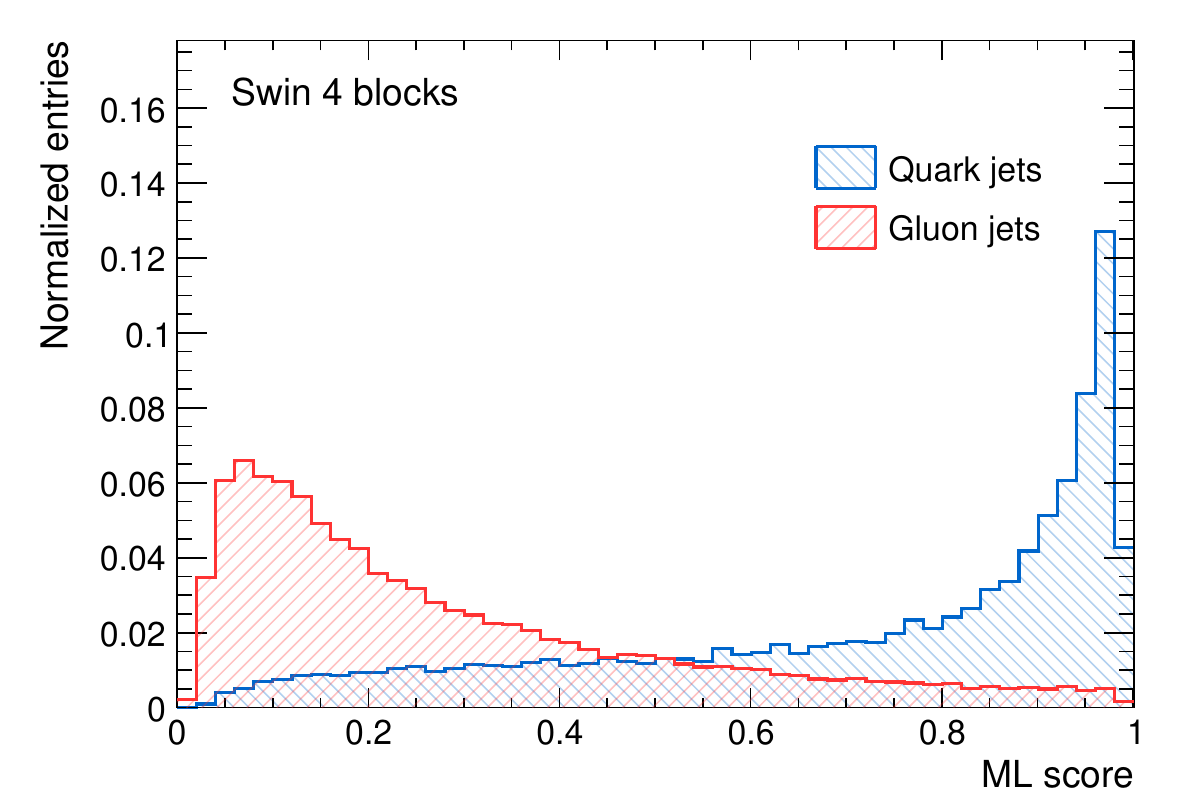}
  \includegraphics[width=7.7cm]{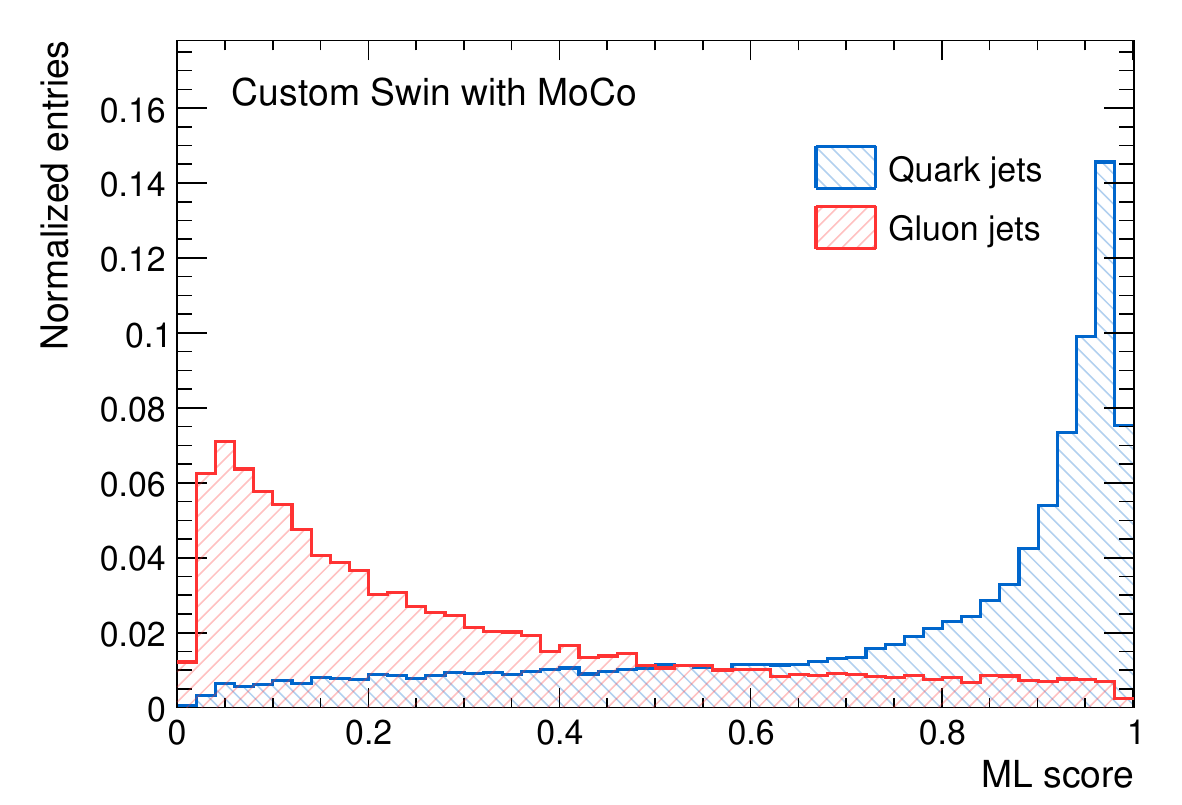}
  \caption{Model output score distributions for quark and gluon jets obtained with CNN (top left), ViT with 0 unfrozen blocks (top right), Swin with 4 unfrozen blocks (bottom left) and custom Swin with MoCo (bottom right). Each distribution is normalized to unit area.}
\label{fig:scoreDists}
\end{figure}

Figure~\ref{fig:scoreDists} shows the corresponding output score distributions for selected models trained on the same 20k dataset. The ML score ranges from gluon-like (0) to quark-like (1).
The ViT model with no fine-tuned blocks shows relatively broad and overlapping quark- and gluon-jet distributions.
In comparison, the 4-block Swin and the custom Swin with MoCo show a clearer separation, with quark and gluon jets concentrated toward high and low scores, respectively.
The CNN also shows clear separation, although its gluon-jet distribution extends further toward intermediate scores than those of the Swin-based models. The small bump structures that appear around an ML score of 0.4 in the CNN are due to the ReLU activation function, which is only used in CNN. The effect of this structure on the results was investigated and found to be negligible.

\begin{figure}[hbt]
  \centering
  \includegraphics[width=7.75cm]{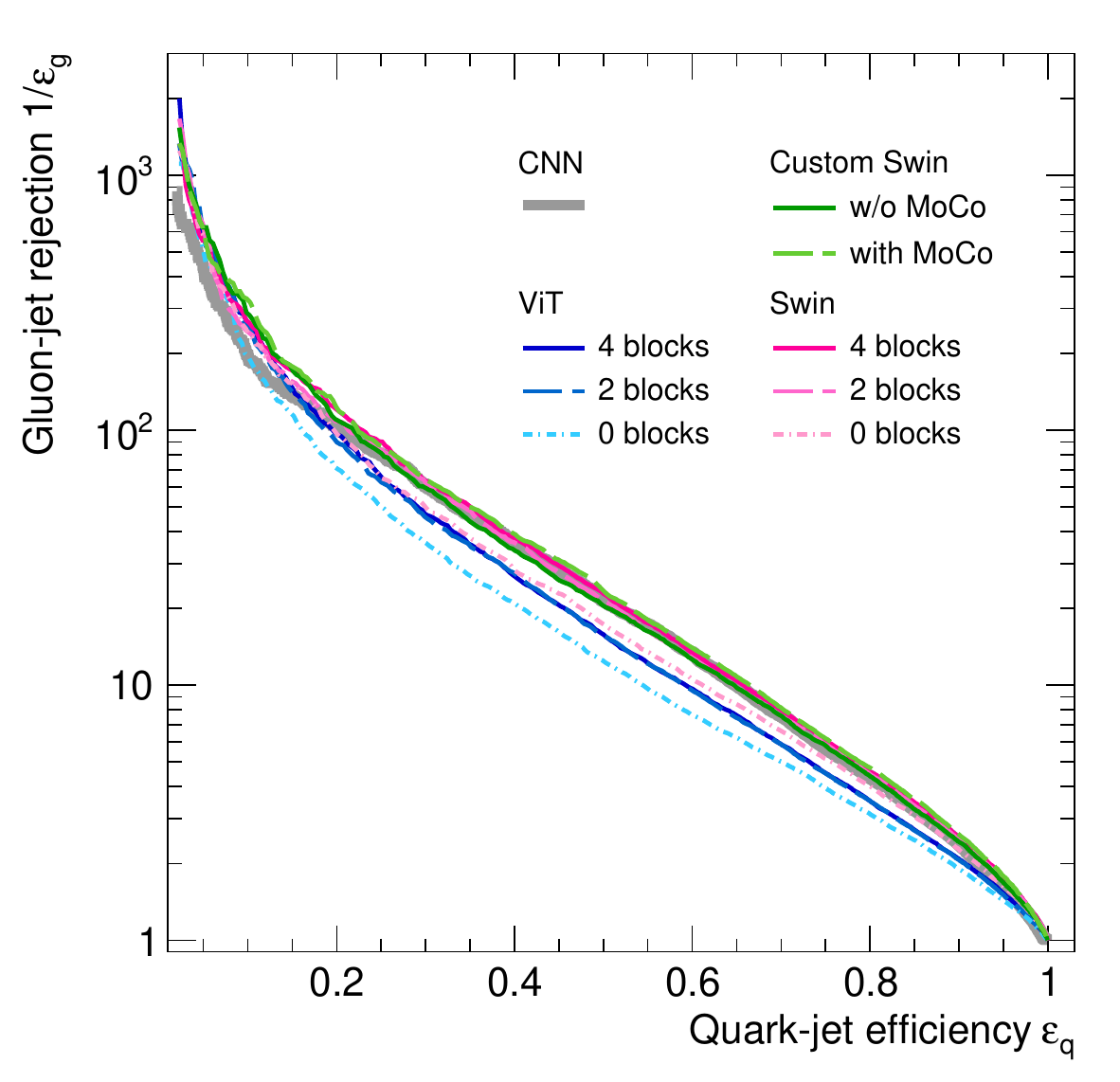}
  \includegraphics[width=7.75cm]{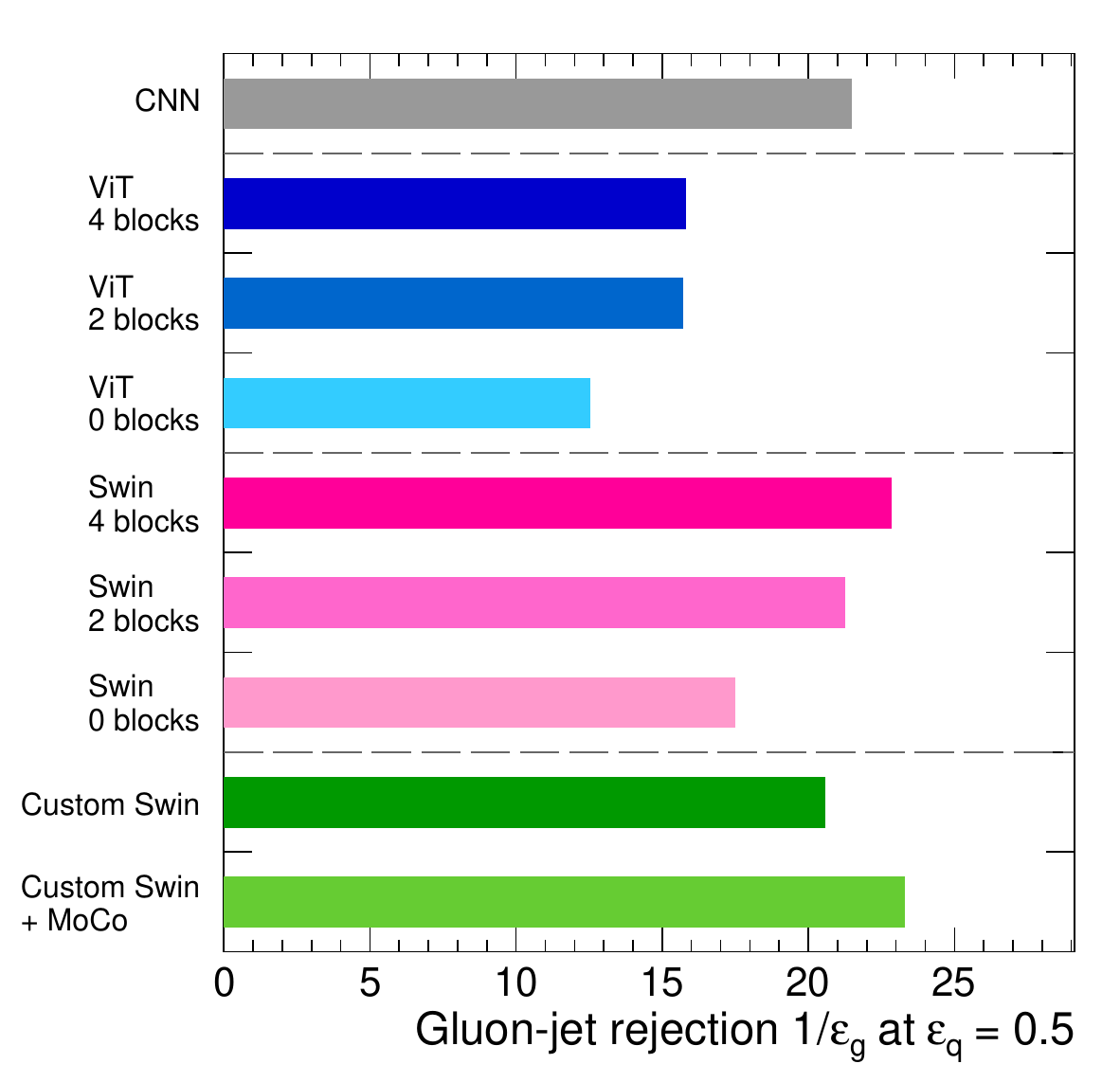}
  \caption{Gluon-jet rejection as a function of quark-jet efficiency (left) and gluon-jet rejection at a quark-jet efficiency of 0.5 (right).}
\label{fig:quarkEff}
\end{figure}

Figure~\ref{fig:quarkEff} shows the gluon-jet rejection as a function of quark-jet efficiency for models trained on the 20k dataset. The quark-jet efficiency ($\epsilon_q$) is defined as the fraction of quark jets correctly identified as quark jets, while the gluon-jet rejection ($1/\epsilon_g$) is the inverse of the fraction of gluon jets misidentified as quark jets.

Swin generally achieves higher gluon-jet rejection than ViT, and its performance improves as more blocks are fine-tuned. For ViT, a substantial improvement is observed from 0 to 2 fine-tuned blocks, while the difference between 2 and 4 blocks is relatively small. The right panel provides a direct comparison at a fixed quark-jet efficiency of $\epsilon_q=0.5$. At this working point, the 4-block Swin and the custom Swin with MoCo give the highest gluon-jet rejection. The improvement with increasing fine-tuning depth is also clearly visible for both ViT and Swin.

\begin{figure}[hbt]
  \centering
  \includegraphics[width=12.0cm]{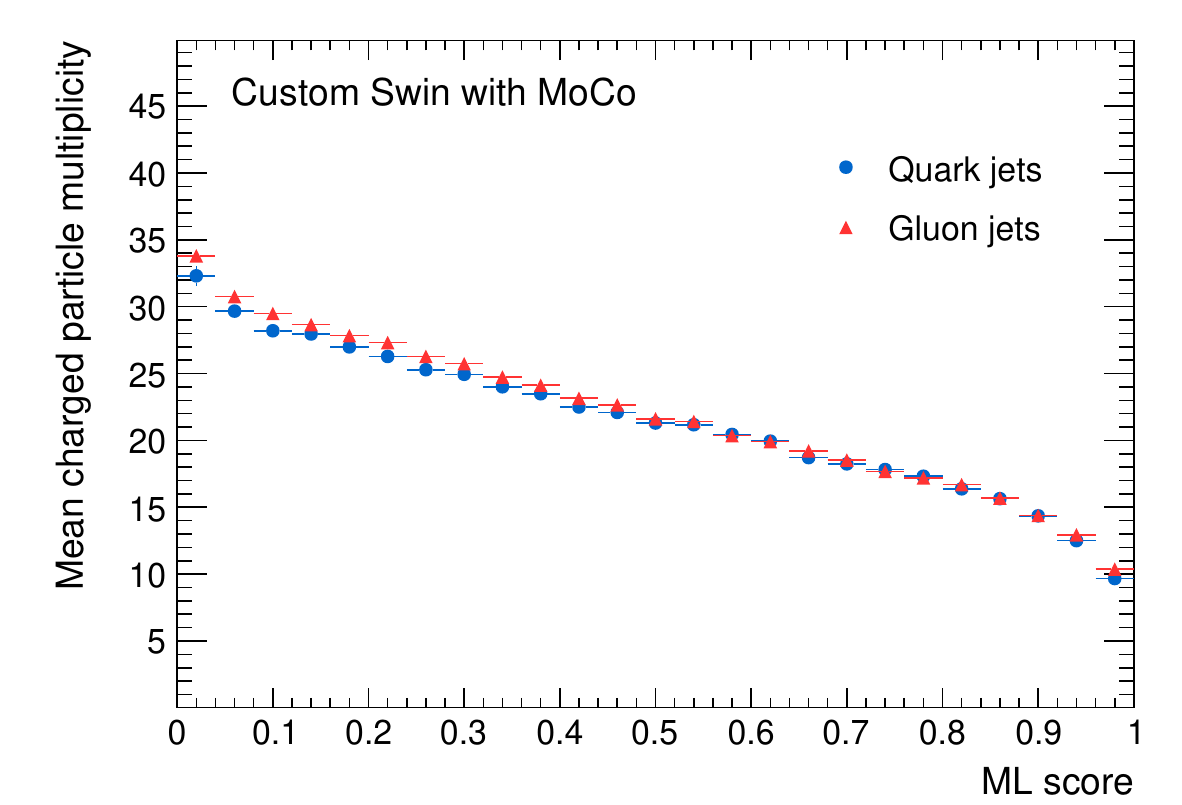}
  \caption{
  Mean charged-particle multiplicity as a function of the model score for quark and gluon jets obtained with the custom Swin with MoCo.}
\label{fig:score_multiplicity}
\end{figure}

To better understand the model response in terms of known jet properties, Figure~\ref{fig:score_multiplicity} shows the mean charged-particle multiplicity as a function of the model score for the custom Swin with MoCo.
The multiplicity decreases as the score moves from gluon-like to quark-like, consistent with the known higher particle multiplicity of gluon jets compared with quark jets.
This correlation indicates that the classifier response reflects charged-particle multiplicity, a conventional observable used for quark--gluon discrimination. A similar trend is observed for the other architectures, as known in Appendix~\ref{sec:score_multiplicity_all}.

\begin{figure}[hbt]
  \centering
  \includegraphics[width=7.75cm]{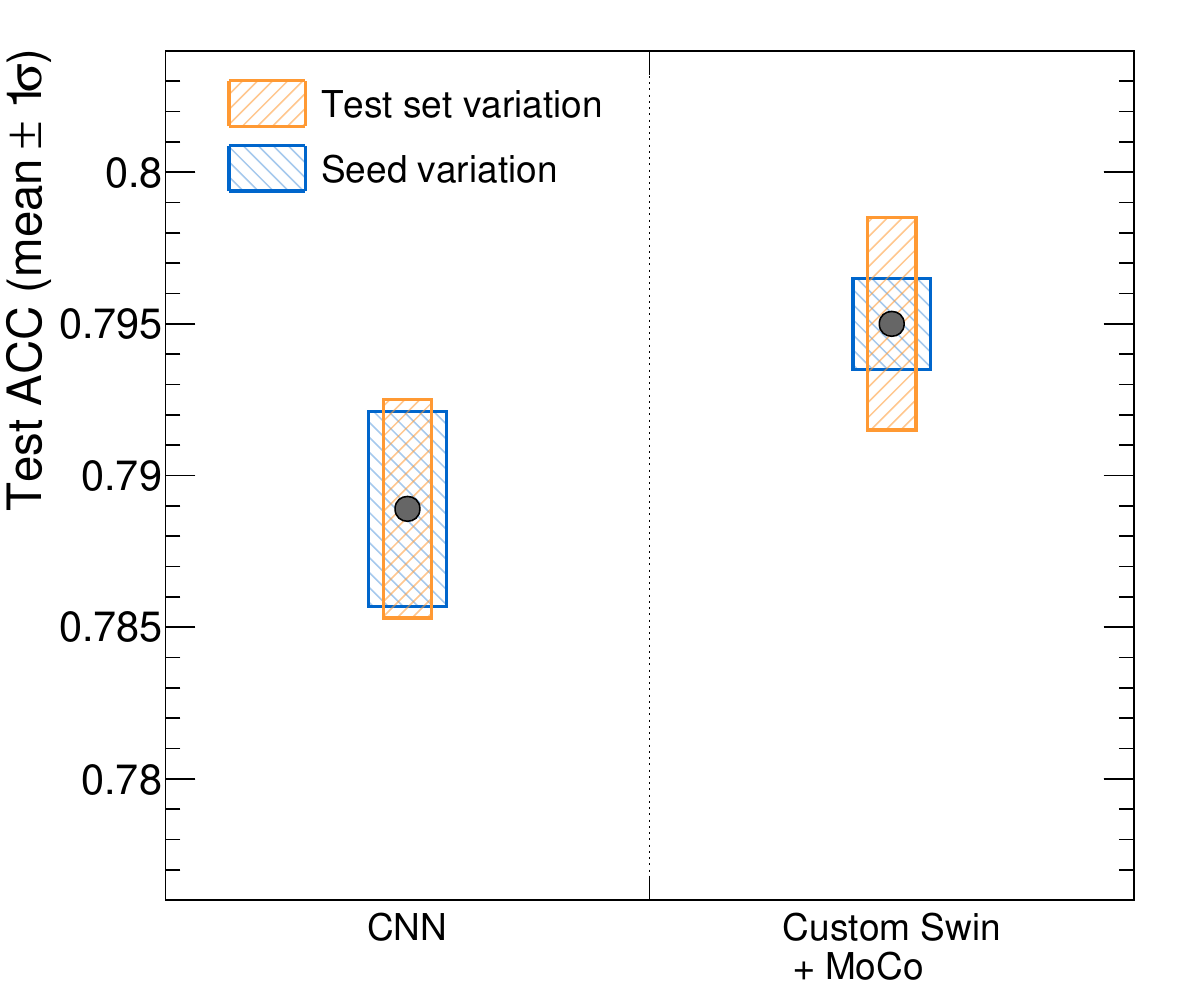}
  \includegraphics[width=7.75cm]{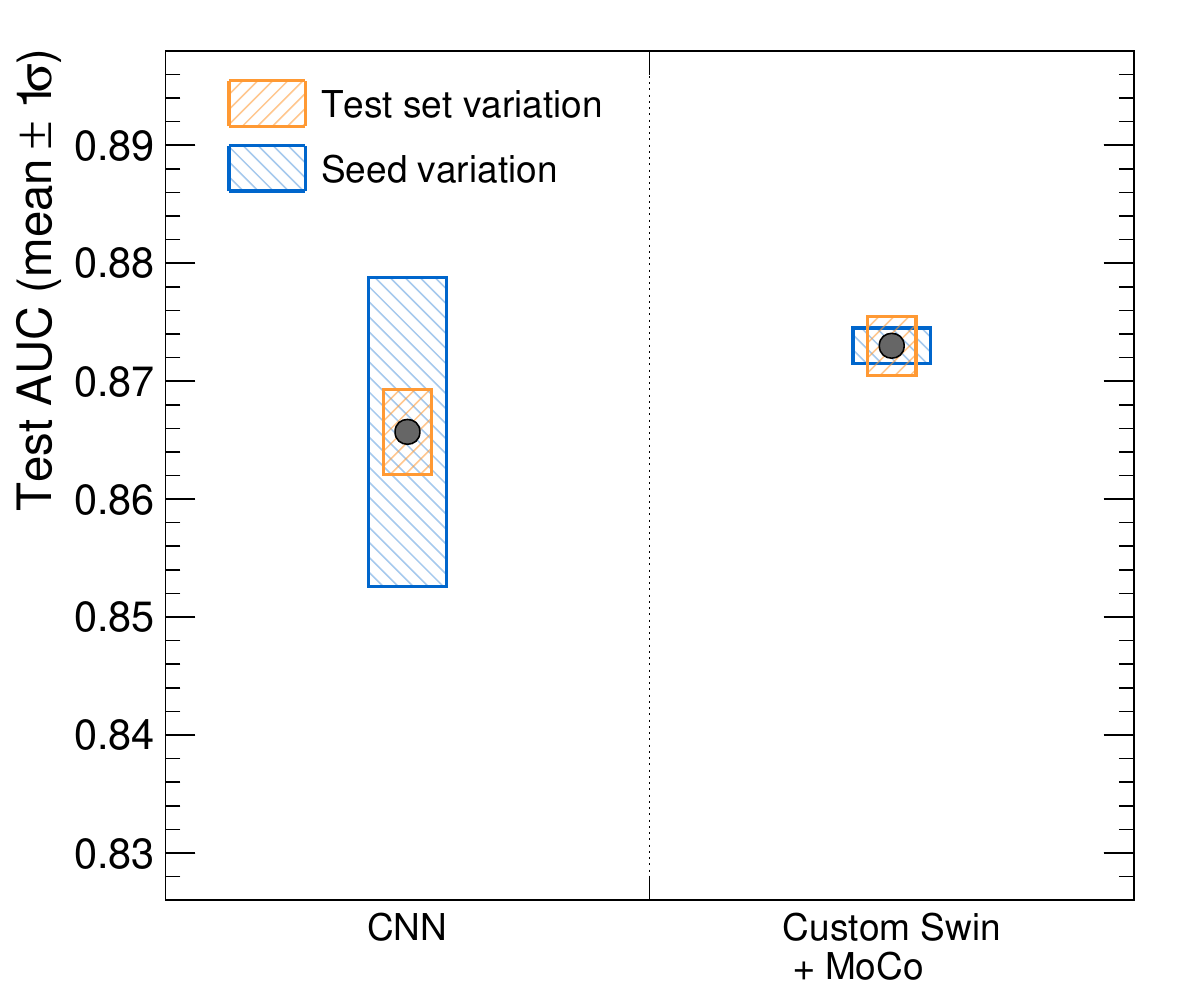}
  \caption{Test ACC (left) and AUC (right) values obtained with CNN and custom Swin with MoCo models, together with their uncertainties due to variations in the test set and random seed used for training. The boxes represent the corresponding mean with one standard deviation. The study was performed using the 20k dataset.}
\label{fig:errEstimation}
\end{figure}

\begin{table}[t]
\centering
\caption{
Test performance and uncertainties for CNN and custom Swin with MoCo on the fixed 40,000-jet test set.
It shows the AUC across ten training seeds, the seed-ensemble AUC with its 95\% bootstrap confidence interval, and the paired comparison between the two ensembles.
}
\label{tab:test_performance_uncertainty}

\small
\setlength{\tabcolsep}{5pt}
\renewcommand{\arraystretch}{1.1}

\begin{tabular}{lccc}
\hline
\multicolumn{4}{c}{\textbf{Test performance}} \\
\hline
\textbf{Model} &
\shortstack{\textbf{AUC}\\\textbf{(mean $\pm$ std)}} &
\shortstack{\textbf{Ensemble}\\\textbf{AUC}} &
\shortstack{\textbf{95\% bootstrap}\\\textbf{CI}} \\
\hline

Custom Swin (MoCo init.) &
$0.8714 \pm 0.0015$ &
0.8767 &
$[0.8734,\ 0.8800]$ \\

CNN &
$0.8529 \pm 0.0129$ &
0.8665 &
$[0.8629,\ 0.8700]$ \\

\hline
\multicolumn{4}{c}{\textbf{Paired ensemble comparison}} \\
\hline
\textbf{Comparison} &
\textbf{$\Delta$AUC} &
\textbf{95\% CI} &
\textbf{$p$ (Holm)} \\
\hline

Custom Swin $-$ CNN &
+0.0102 &
$[+0.0083,\ +0.0120]$ &
0.0011 \\

\hline
\end{tabular}

\end{table}

To evaluate the stability of the model, CNN and custom Swin with MoCo were selected for additional tests with variations in the training seed and test set. For seed variation, each model was trained ten times with different random seeds, while for test set variation, a single trained model was evaluated on ten independent test sets.
The resulting variations in ACC and AUC were used to estimate the corresponding uncertainties.
Figure~\ref{fig:errEstimation} compares these uncertainties for the two models.
The test-set uncertainties are similar for the two models, whereas the CNN shows a larger dependence on the training seed, particularly for the AUC.

For a more quantitative comparison, a seed ensemble was constructed by averaging the predicted probabilities of the ten independently trained models on the fixed 40,000-jet test set.
The 95\% confidence intervals of the ensemble AUCs were estimated using 10,000 stratified bootstrap resamples.
The difference between the two ensemble AUCs was evaluated using a paired bootstrap and the corresponding $p$-value was obtained from a two-sided paired permutation test with 10,000 permutations.
A Holm correction was applied for multiple comparisons. The results are summarized in Table~\ref{tab:test_performance_uncertainty}.

The custom Swin with MoCo shows a higher mean AUC and smaller seed-to-seed variation than the CNN.
The ensemble comparison further shows a positive AUC difference with a confidence interval excluding zero and a statistically significant Holm-corrected $p$-value.
Together with the seed-variation results, this indicates that the performance advantage of the custom Swin with MoCo is consistent across training seeds and statistically significant on the fixed test sample.

\subsubsection{Effect of Block-wise Fine-tuning}

\begin{figure}[hbt]
  \centering
  \includegraphics[width=7.75cm]{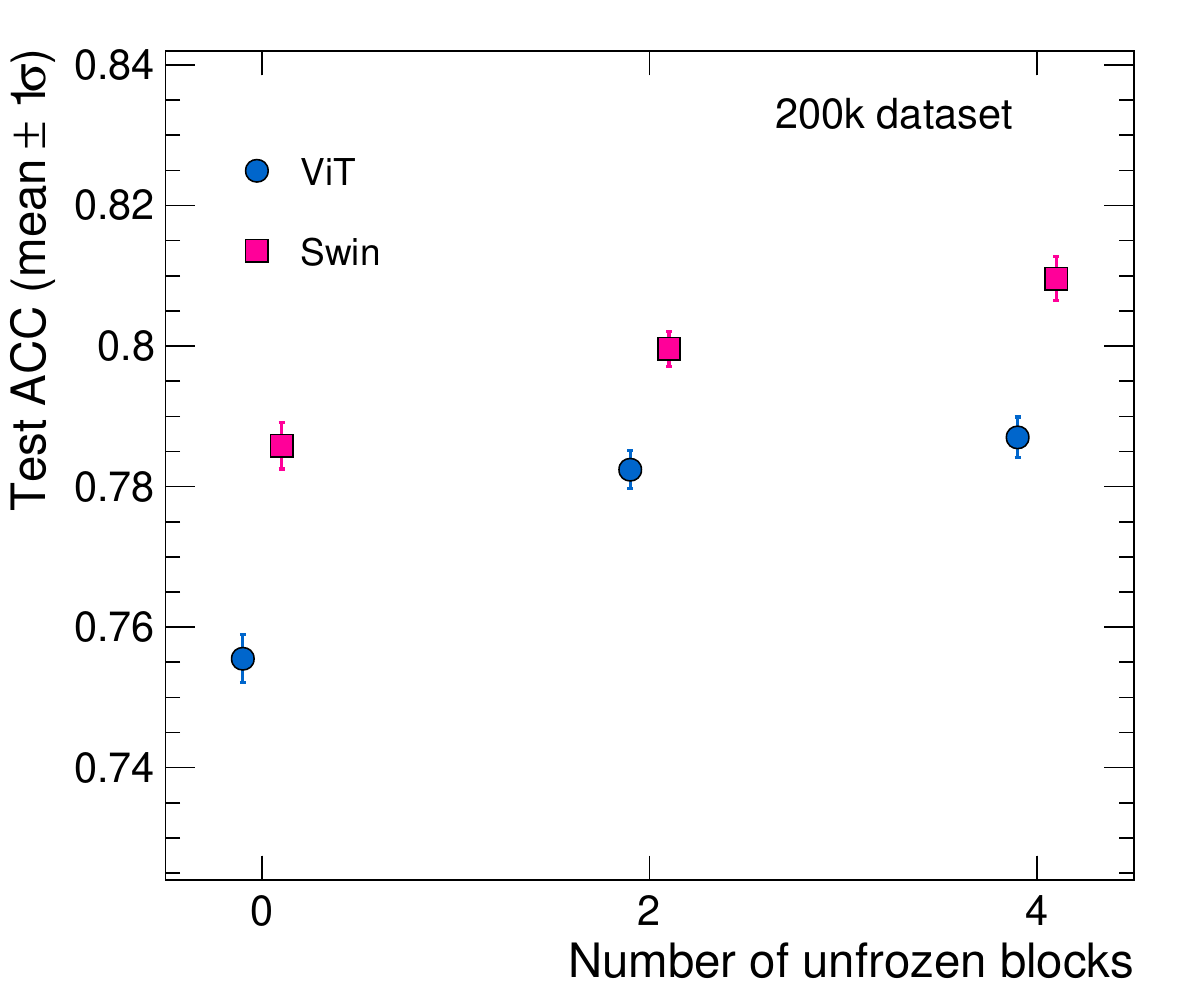}
  \includegraphics[width=7.75cm]{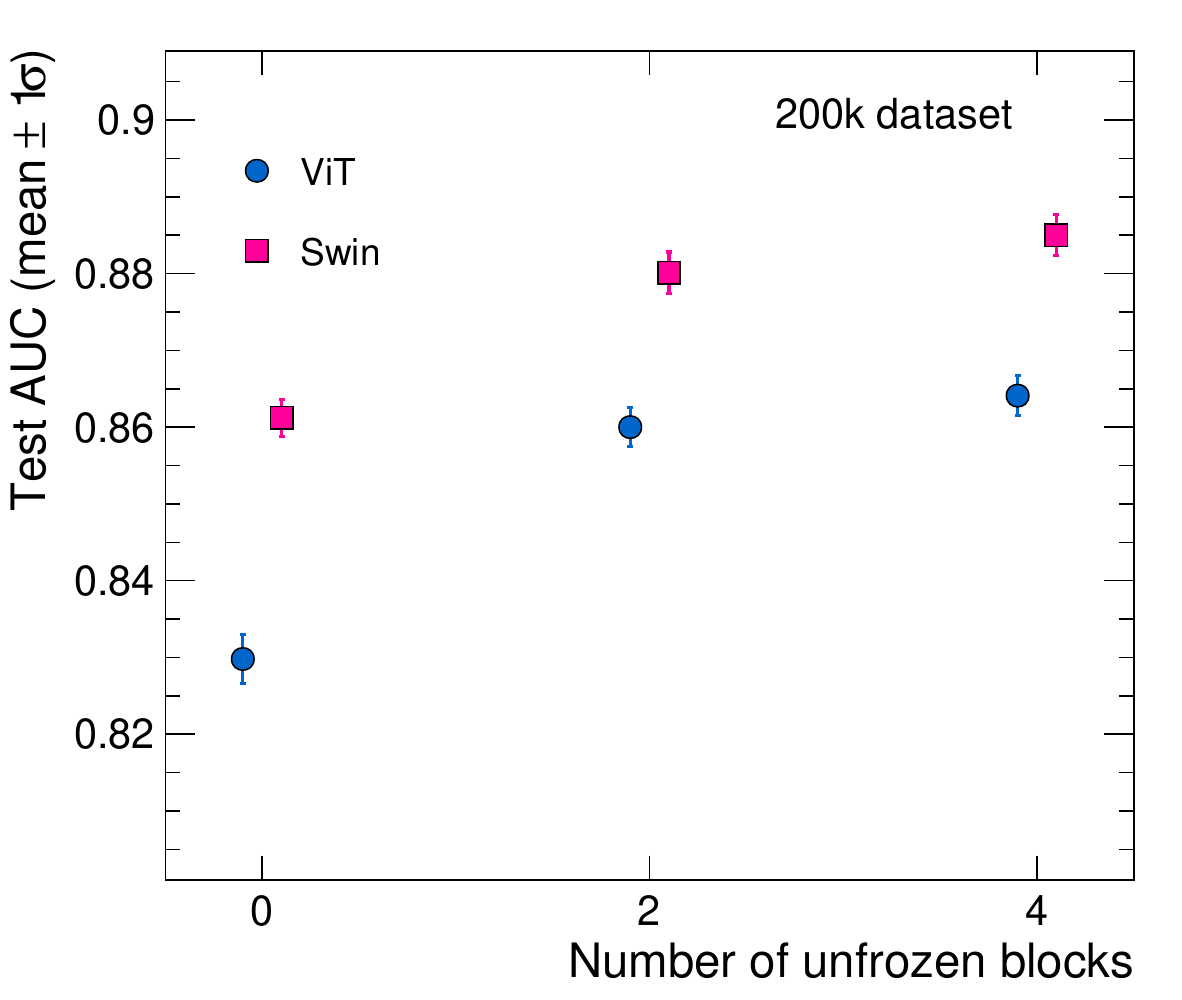}
  \caption{Effect of fine-tuning depth on ACC and AUC for ViT and Swin models.
Results are shown for ViT (blue circle) and Swin (red square) datasets, with error bars indicating one standard deviation obtained from the variation across the test sets.
The data points are slightly shifted along the $x$-axis (number of unfrozen blocks) for better visibility.}
\label{fig:depth}
\end{figure}

Figure~\ref{fig:depth} shows the same ACC and AUC values as Figure~\ref{fig:all_test}, rearranged for the 200k dataset with fine-tuning depths of 0, 2 and 4.
As shown in Figure~\ref{fig:depth}, performance generally improves as the fine-tuning depth increases, with models using deeper fine-tuning consistently outperforming those with fewer fine-tuned blocks across most configurations.
Both ViT and Swin show substantial performance improvements when increasing the fine-tuning depth from 0 to 2 blocks, but only marginal gains when further increasing it from 2 to 4 blocks.
This behavior suggests that while deeper fine-tuning contributes to improved classification performance, its benefits become limited under a moderate-scale dataset setting, where increasing the number of fine-tuned blocks may not lead to significant performance gains.

\subsubsection{Impact of MoCo Self-supervised Pretraining}

Figure~\ref{fig:all_test} also includes the performance of the custom Swin models with MoCo self-supervised pretraining, indicated by the markers labeled ``Custom Swin + MoCo''.
The models initialized with MoCo pretraining generally show improved performance compared with their non-pretrained counterparts. As discussed in Section 3.4, this improvement may be related to the contrastive learning objective of MoCo, which provides a different initialization by learning representations from the jet images before supervised training. These results suggest that self-supervised pretraining can provide a useful initialization for the custom Swin model.

\subsubsection{Effect of dataset size on model performance}

\begin{figure}[htb]
  \centering
  \includegraphics[width=7.75cm]{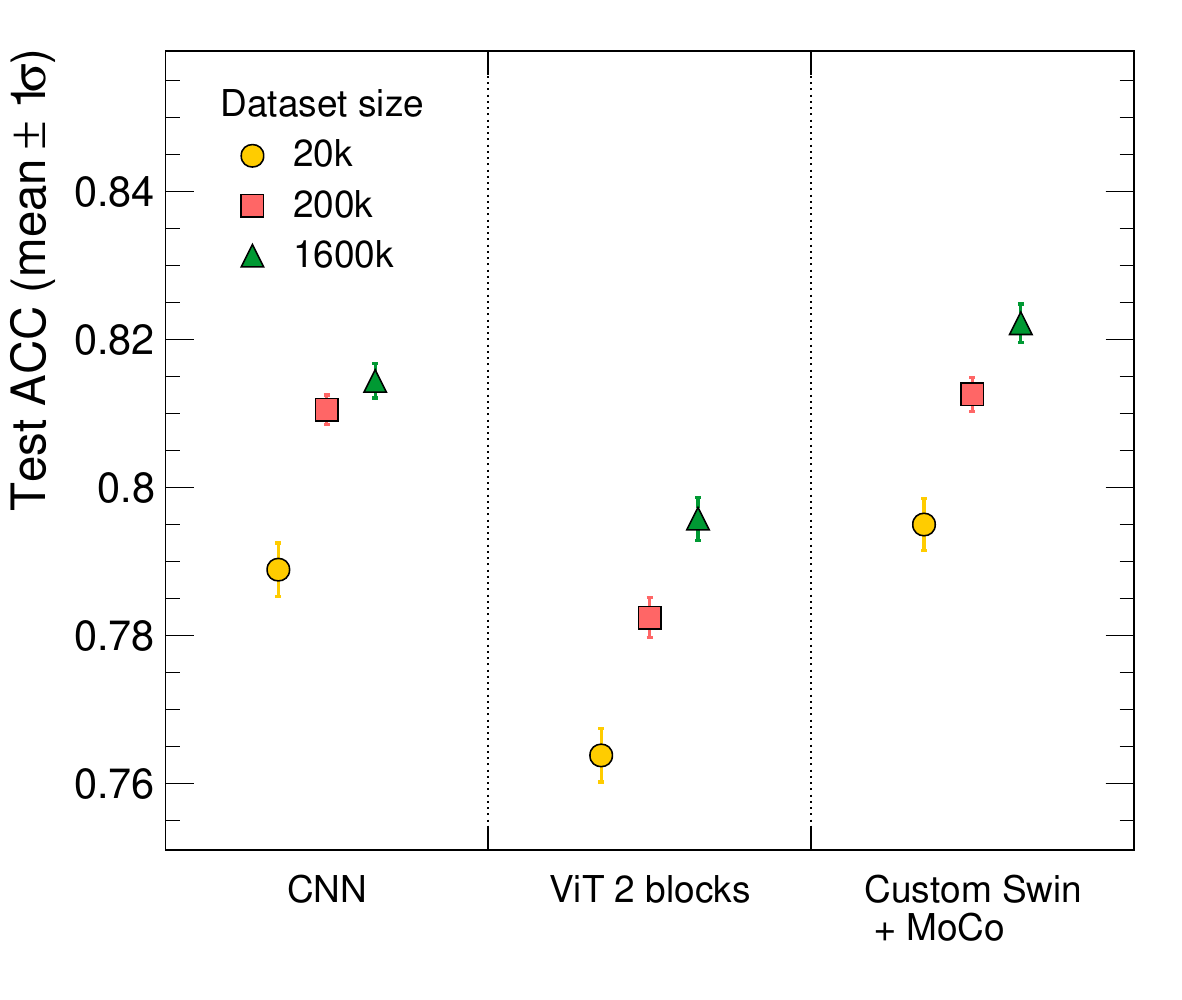}
  \includegraphics[width=7.75cm]{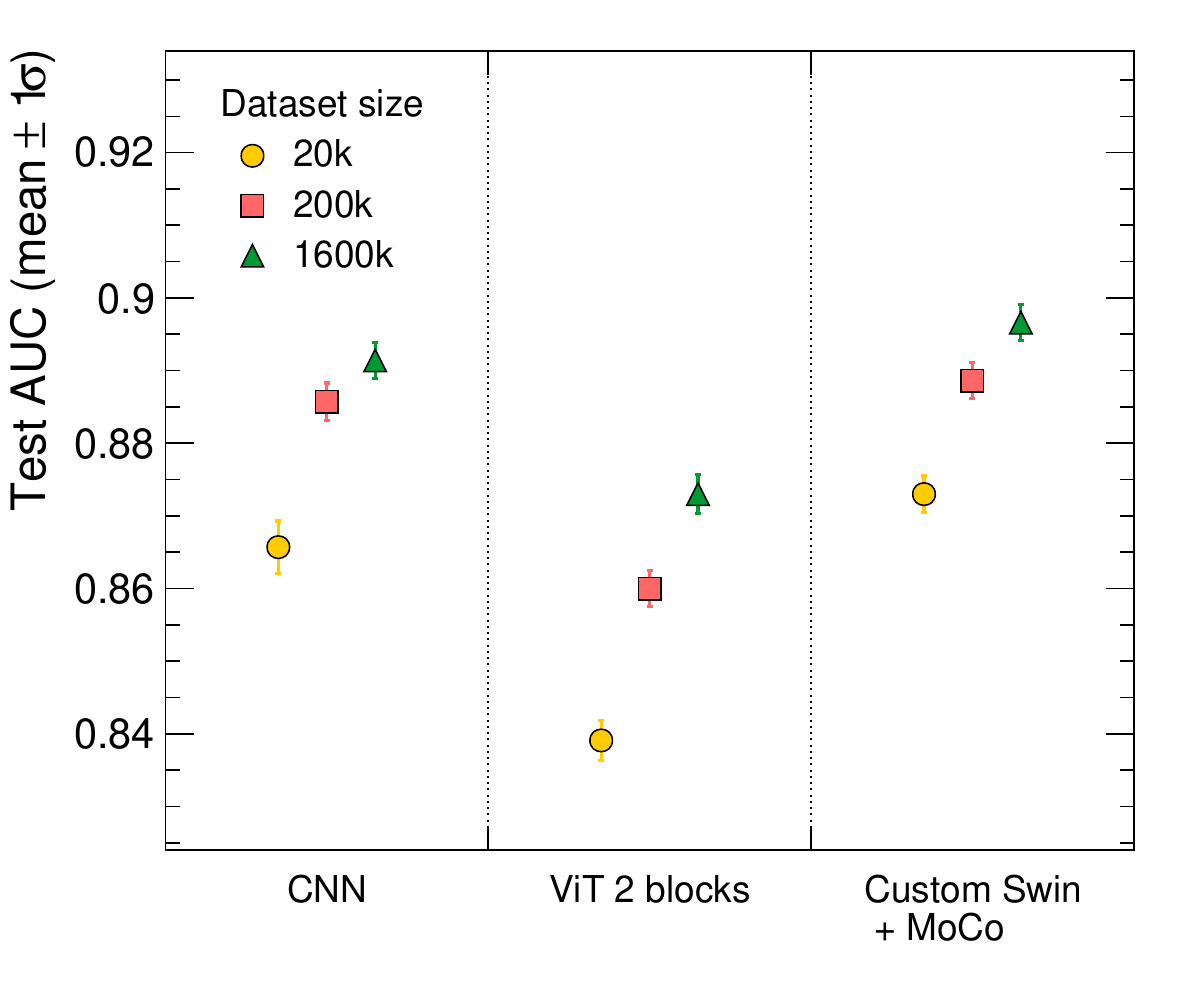}
  \caption{Test ACC (left) and AUC (right) values obtained with the 1.6M dataset together with those from the 20k and 200k datasets shown in Figure~\ref{fig:all_test} for CNN, ViT with a fine-tuning 2 blocks, and a custom Swin incorporating MoCo. Error bars represent one standard deviation obtained from the test set variation.}
\label{fig:dataset1600}
\end{figure}

Given the observation in Figure~\ref{fig:all_test} that larger datasets tend to improve model performance, we further explore the effect of increasing dataset size. For this study, CNN, ViT with a fine-tuning 2 blocks and a custom Swin with MoCo were trained on the 1.6M dataset. Figure~\ref{fig:dataset1600} shows the test ACC and AUC values obtained with the 1.6M dataset, compared with those from the 20k and 200k datasets shown in Figure~\ref{fig:all_test}. While performance continues to improve for all three models, the gain from 200k to 1.6M is smaller than that from 20k to 200k, suggesting a tendency toward smaller marginal gains with increasing dataset size. In addition, the improvements from 200k to 1.6M for the ViT fine-tuning 2 blocks and the custom Swin with MoCo are larger than those for the CNN, indicating that these models may benefit more from further increases in dataset size.

\FloatBarrier

\subsection{Training Dynamics and Efficiency}

\subsubsection{Epoch-based Validation Analysis (Training Dynamics)}

\begin{figure}[h]
  \centering
  \includegraphics[width=7.2cm]{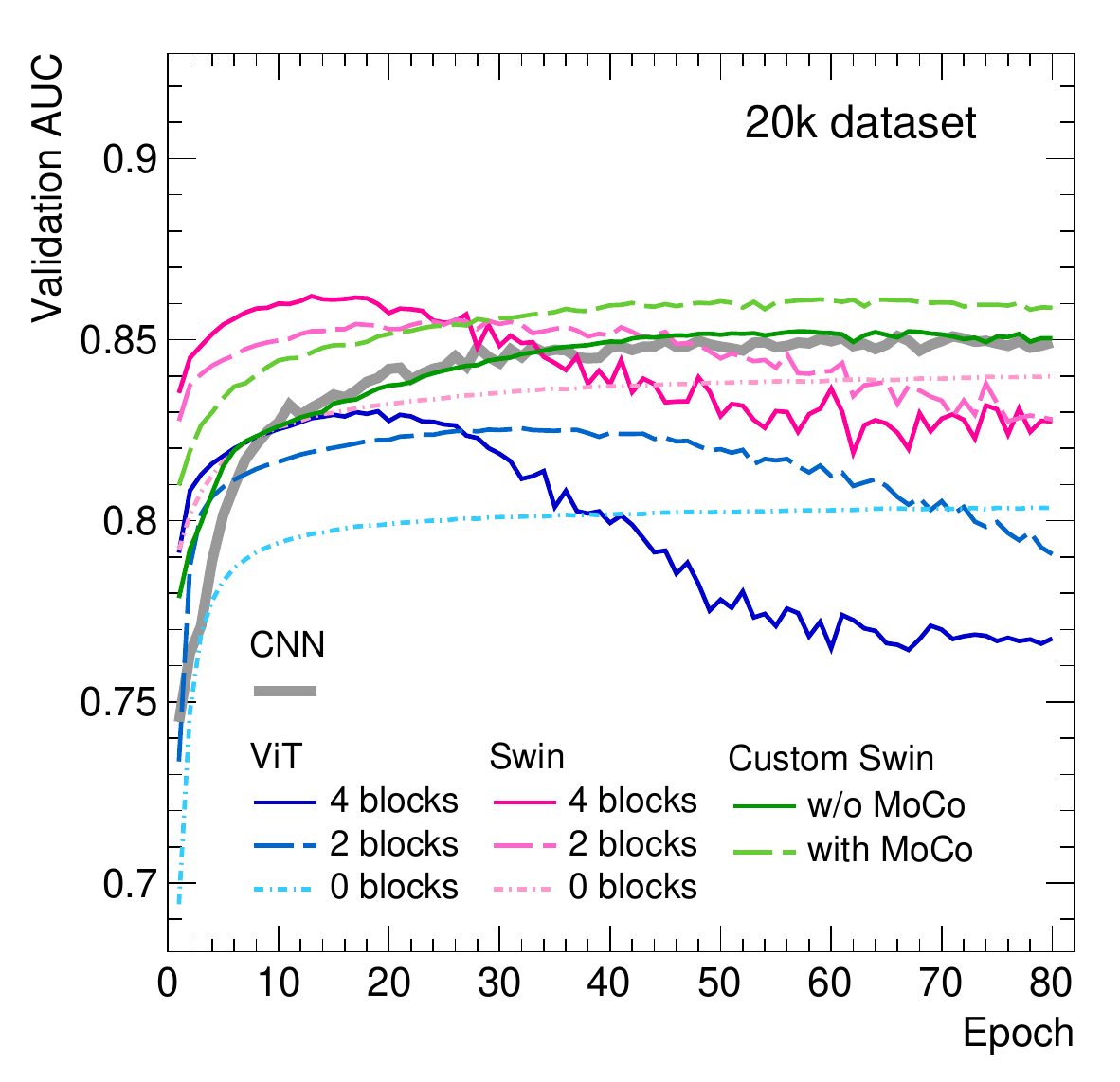}
  \includegraphics[width=7.2cm]{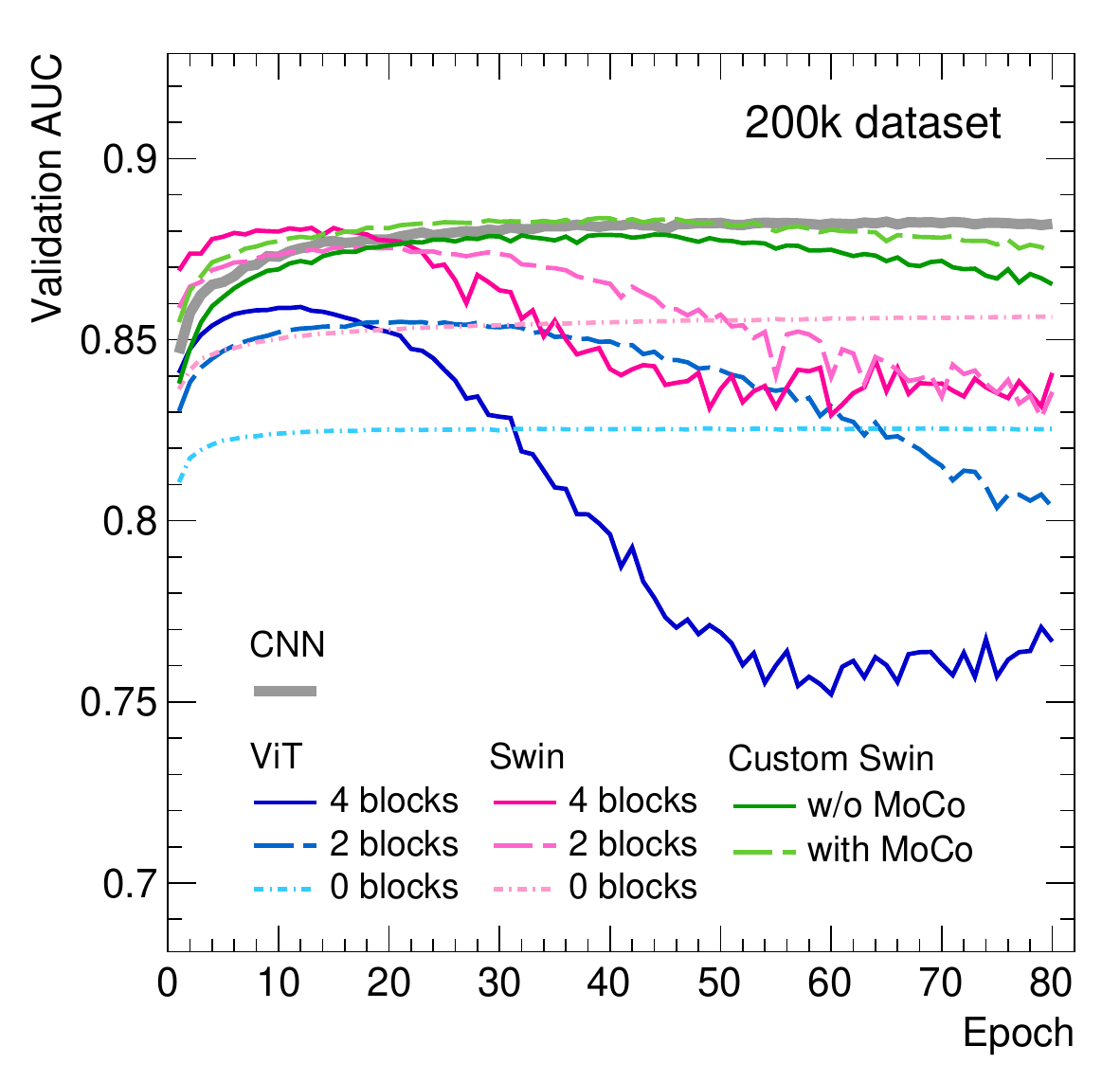}
  \caption{Validation AUC as a function of training epochs for different model architectures and fine-tuning configurations obtained from the 20k (left) and 200k (right) datasets. The figure illustrates differences in learning dynamics across fine-tuning depths and dataset sizes.}
\label{fig:val_vs_epoch}
\end{figure}

Figure~\ref{fig:val_vs_epoch} shows the validation AUC as a function of training epochs for different model architectures and fine-tuning configurations. The performance comparisons in Figures ~\ref{fig:all_test}-~\ref{fig:dataset1600} were obtained using the checkpoint at the epoch where each model configuration achieved its highest validation AUC.
Across all dataset sizes, configurations with deeper fine-tuning (i.e., more unfrozen backbone blocks) tend to reach their maximum validation AUC at earlier stages of training, followed by a subsequent decline in performance.
This behavior is consistent with reduced optimization stability when a large number of parameters are updated simultaneously. When many backbone parameters are unfrozen relative to the effective training signal provided by the data, the optimization process can become more sensitive to sample-level fluctuations, resulting in earlier performance saturation and subsequent degradation in validation performance.
Increasing the dataset size improves the achievable maximum validation AUC, but does not eliminate this behavior in our study.
In contrast, configurations employing minimal fine-tuning, such as the 0 block setting, exhibit smoother learning dynamics and more gradual improvements in validation performance.
Overall, these results show the need to balance the number of trainable parameters and the available training statistics when selecting fine-tuning strategies. 

\FloatBarrier
\subsubsection{Time-based Validation Analysis (Computational Efficiency)}

\begin{figure}[htb]
  \centering
  \includegraphics[width=7.2cm]{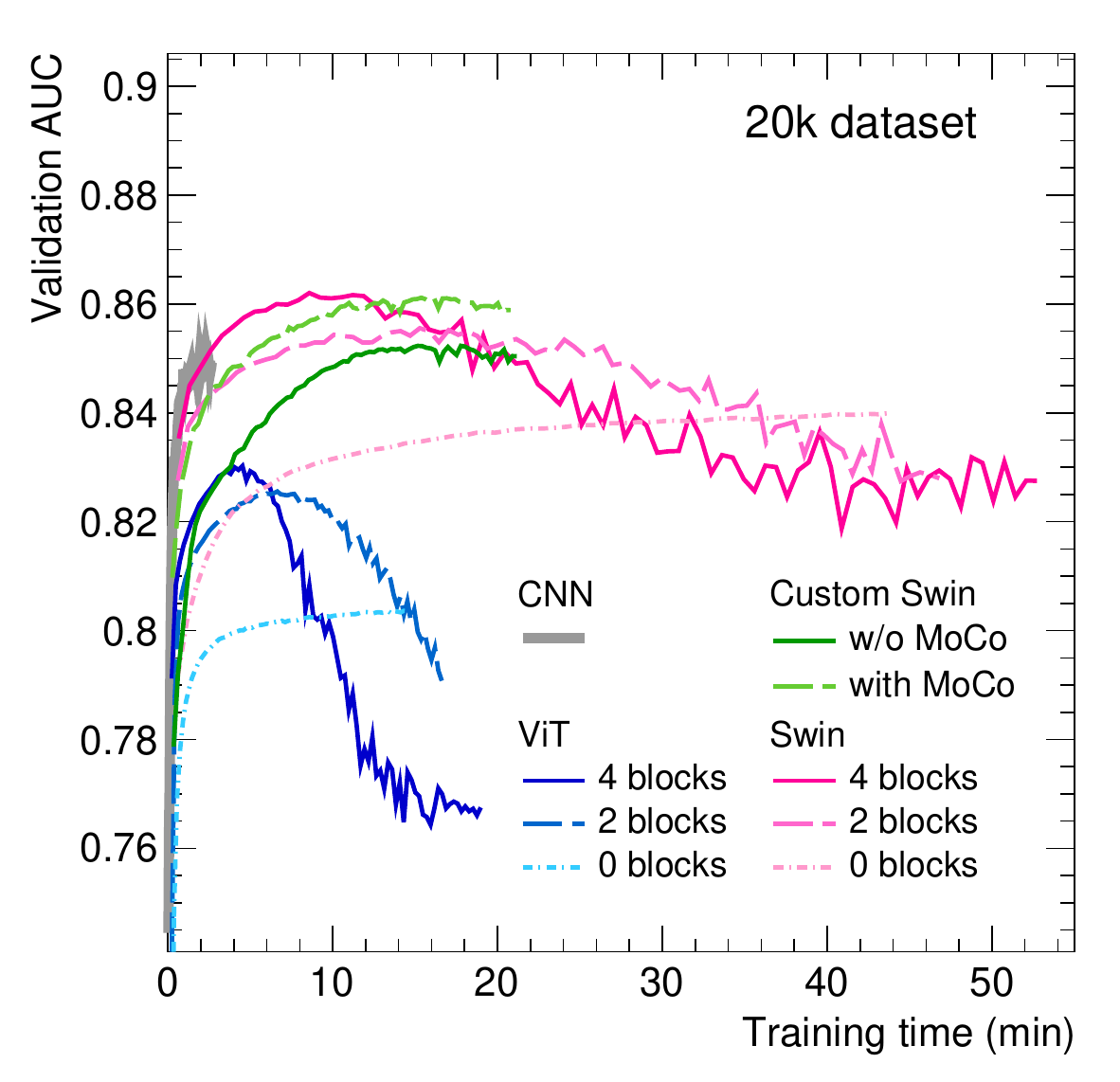}
  \includegraphics[width=7.2cm]{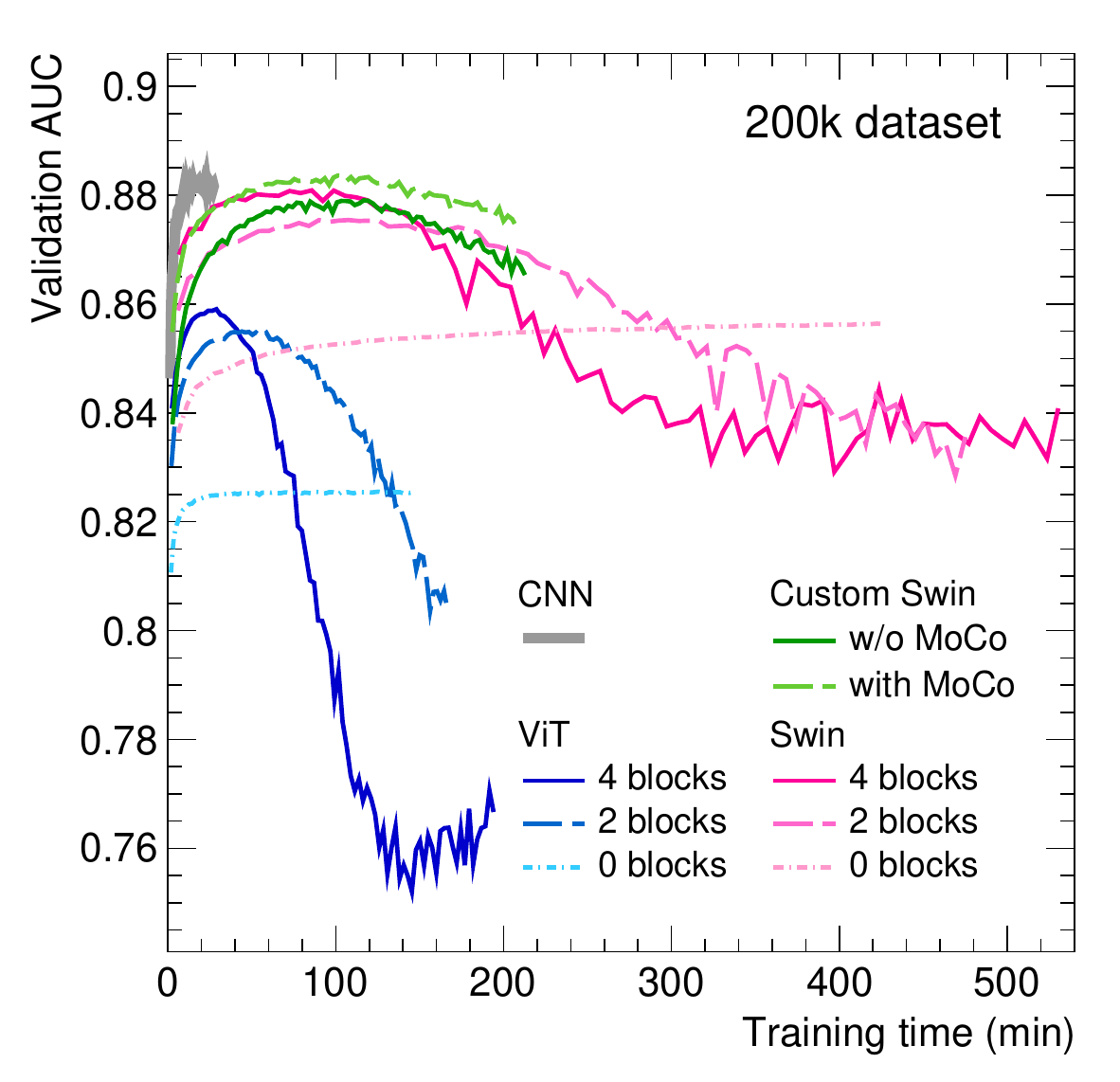}
  \caption{Validation AUC as a function of training time for different model architectures and fine-tuning configurations obtained from the 20k (left) and 200k (right) datasets.}
\label{fig:val vs time}
\end{figure}

Figure~\ref{fig:val vs time} shows the validation AUC as a function of the actual training time for different model architectures and fine-tuning configurations. In contrast to the epoch-based comparison, this representation provides a more direct view of computational efficiency and convergence behavior.
Across all dataset sizes, most model configurations reach near-optimal validation performance within a broadly similar training time window. Although small differences in convergence speed are visible, the overall time scale required to approach peak performance does not vary significantly between architectures or fine-tuning depths. In particular, the CNN and the ViT configuration with four unfrozen backbone blocks tend to reach their maximum validation AUC slightly earlier than the other models.
As the dataset size increases, the achievable maximum validation AUC generally improves, while the time required to reach their maximum validation AUC also becomes longer due to the increased computational cost per epoch. Overall, these results indicate that most model configurations reach near-optimal validation performance within a comparable training time window.

\begin{table}[t]
\centering
\caption{Model complexity and inference efficiency of the evaluated models. Params denotes the number of model parameters in millions and GMACs the multiply-accumulate operations per jet.
Latency is the median forward-pass time at batch size $B=1$. GPU and CPU throughput (TP) are measured in jets/s at $B=128$ and $B=32$,
respectively.}
\label{tab:model_efficiency}
\begin{tabular}{lcccccc}
\hline
\textbf{Model} & 
\makecell{\textbf{Params} \\ \textbf{(M)}} &
\makecell{\textbf{GMACs} \\ \textbf{(/jet)}} &
\makecell{\textbf{Latency} \\ \textbf{(ms)}} &
\makecell{\textbf{GPU throughput} \\ \textbf{(jets/s)}} &
\makecell{\textbf{CPU throughput} \\ \textbf{(jets/s)}}\\
\hline
CNN &   1.42 & 0.054 & 0.20 & 38697.0 & 1790.2 \\

ViT &  5.52 & 1.080 & 1.43 & 3447.3 & 156.6 \\

Swin &  27.52 & 4.509 & 2.89 & 750.7 & 37.0 \\

Custom Swin  & 3.82 & 0.542 & 1.82 & 4626.1 & 245.8 \\

\hline
\end{tabular}
\end{table}

Table~\ref{tab:model_efficiency} compares the model complexity and inference efficiency of the evaluated architectures.
The CNN is the most computationally efficient model, with 1.42M parameters and 0.054 GMACs per jet. It also shows the shortest $B=1$ latency and the highest GPU and CPU throughput.
In contrast, Swin has the largest number of parameters and the highest computational cost, resulting in the longest latency and lowest throughput among the models considered.

The custom Swin substantially reduces the model size and computational cost relative to the standard Swin, with 3.82M parameters and 0.542 GMACs per jet. It also has fewer parameters and lower computational cost than ViT and achieves higher GPU and CPU throughput, although its $B=1$ latency is slightly longer than that of ViT. These results show that the custom Swin substantially reduces the computational cost of the standard Swin architecture while retaining the hierarchical representation of the Swin Transformer.

\FloatBarrier
\section{Discussion and Outlook}

In this work, we compared CNN, ViT, and Swin-based models for quark--gluon classification using a common jet-image representation. The study focused on how the classification performance changes with model architecture, fine-tuning depth, training sample size, and self-supervised pretraining.

Across the configurations considered, CNN and Swin-based models generally show better classification performance than ViT. Since both CNN and Swin incorporate local feature processing in their architectures, this behavior may be related to the importance of local structures in jet images. The hierarchical structure of Swin may also provide a useful way of combining information at different spatial scales. However, the present comparison does not isolate which particular features or spatial scales are responsible for the observed differences between the models.

Some information about the classifier response can be obtained from its correlation with charged-particle multiplicity. For all models, jets with more gluon-like scores tend to have larger charged-particle multiplicities, consistent with the known difference between quark and gluon jets. This suggests that the models make use, at least in part, of conventional quark--gluon discriminating information. Further studies using other jet observables or interpretability methods would be needed to understand what additional information is learned from the spatial structure of the jet images.

The fine-tuning studies show a moderate dependence on the number of unfrozen Transformer blocks. In general, moving from zero to two unfrozen blocks improves the performance, while the difference between two and four blocks is smaller. The validation curves also show that models with more unfrozen blocks tend to reach their best performance earlier and, in some cases, decrease afterwards. This behavior is more visible for the smaller training sample and becomes less pronounced as the available statistics increase. These observations suggest that the choice of fine-tuning depth should be made together with the available training statistics rather than assuming that deeper fine-tuning will always be beneficial.

A similar dependence is observed when the size of the training sample is increased. Performance improves from 20k to 200k and further to 1.6M jets for the models studied, although the gain becomes smaller at larger sample sizes. The relative improvement with increasing statistics is not identical among the architectures, suggesting that their dependence on the amount of training data may also differ.

The custom Swin was introduced to examine whether a smaller architecture adapted to the native $72\times72$ jet-image resolution could retain the performance of the larger Transformer models at lower computational cost. In the configurations studied here, its classification performance is comparable to that of the larger Swin models, while requiring substantially fewer parameters and operations. MoCo pretraining provides an additional, although relatively modest, improvement.

The present study has several limitations. The results are based on truth-level simulated jets and do not include detector response, pile-up, generator dependence, or differences between simulation and experimental data. These effects may modify both the jet-image representation and the features used for classification. The results should therefore be regarded primarily as a comparison of architectures and training strategies under a controlled simulation setup.

An important next step is to repeat the comparison with alternative event generators and detector-level simulations. For applications to experimental data, the domain shift between simulation and data will also need to be addressed. Since quark and gluon labels are not directly available in collision data, approaches based on domain adaptation or other weakly- or unsupervised strategies may be useful in this context.

Finally, this study is restricted to image-based representations. Constituent-based and graph-based approaches retain information that may be partly lost when jets are projected onto a fixed image grid. A comparison under common datasets and training conditions would therefore be useful for understanding how much of the behavior observed here is specific to the image representation.

Overall, the results illustrate how the performance of image-based quark--gluon classification depends on several interconnected choices, including architecture, fine-tuning strategy, training statistics, and pretraining. The present study provides a controlled comparison of these effects, while studies with more realistic simulations and experimental data will be needed to determine how these observations translate to practical jet-tagging applications.

\section*{Acknowledgments}
This work was supported by the National Research Foundation of Korea (NRF) under Grant Nos. RS-2008-NR007226 and RS-2025-00559626.
The authors acknowledge the open dataset provided by Zenodo and the computational resources of the Inha Physics AI Laboratory.

\section*{Data availability}
The datasets used in this study were generated using Monte Carlo 
simulations with the configurations described in the manuscript. 
The generated datasets and additional information required to 
reproduce the results are available from the corresponding author 
upon reasonable request.

\appendix
\section{Appendix}
\subsection{Detailed Numerical Results}
Detailed numerical results for all models are provided in
Table~\ref{tab:appendix_results}.

\begin{table}[t]
\centering
\caption{Test performance of all models. Results are reported as mean $\pm$ one standard deviation ($1\sigma$) over multiple runs.}
\label{tab:appendix_results}
\resizebox{\linewidth}{!}{%
    \begin{tabular}{lccc}
    \toprule
    \textbf{Model} &
    \textbf{Test AUC (mean $\pm$ 1$\sigma$)} &
    \textbf{Test ACC (mean $\pm$ 1$\sigma$)} &
    \textbf{Test Loss (mean $\pm$ 1$\sigma$)} \\
    \midrule
    CNN (20k) & 0.8657 $\pm$ 0.0036 & 0.7889 $\pm$ 0.0036 & 0.6752 $\pm$ 0.0855 \\
    CNN (200k) & 0.8857 $\pm$ 0.0026 & 0.8105 $\pm$ 0.0020 & 0.4595 $\pm$ 0.0205 \\
    CNN (1.6M) & 0.8914 $\pm$ 0.0025 & 0.8144 $\pm$ 0.0023 & 0.4693 $\pm$ 0.0222 \\\hline
    
    ViT block0 (20k) & 0.8235 $\pm$ 0.0030 & 0.7473 $\pm$ 0.0037 & 0.5230 $\pm$ 0.0042 \\
    ViT block0 (200k) & 0.8298 $\pm$ 0.0032 & 0.7555 $\pm$ 0.0034 & 0.5089 $\pm$ 0.0041 \\
    ViT block2 (20k) & 0.8391 $\pm$ 0.0028 & 0.7638 $\pm$ 0.0036 & 0.4978 $\pm$ 0.0042 \\
    ViT block2 (200k) & 0.8600 $\pm$ 0.0025 & 0.7824 $\pm$ 0.0027 & 0.4661 $\pm$ 0.0040 \\
    ViT block2 (1.6M) & 0.8730 $\pm$ 0.0027 & 0.7958 $\pm$ 0.0029 & 0.4466 $\pm$ 0.0046 \\
    ViT block4 (20k) & 0.8405 $\pm$ 0.0026 & 0.7634 $\pm$ 0.0022 & 0.4996 $\pm$ 0.0040 \\
    ViT block4 (200k) & 0.8641 $\pm$ 0.0026 & 0.7870 $\pm$ 0.0029 & 0.4600 $\pm$ 0.0044 \\\hline
    
    Swin block0 (20k) & 0.8518 $\pm$ 0.0022 & 0.7535 $\pm$ 0.0033 & 0.5150 $\pm$ 0.0038 \\
    Swin block0 (200k) & 0.8612 $\pm$ 0.0024 & 0.7858 $\pm$ 0.0033 & 0.4659 $\pm$ 0.0036 \\
    Swin block2 (20k) & 0.8656 $\pm$ 0.0026 & 0.7846 $\pm$ 0.0033 & 0.4675 $\pm$ 0.0047 \\
    Swin block2 (200k) & 0.8801 $\pm$ 0.0027 & 0.7996 $\pm$ 0.0025 & 0.4464 $\pm$ 0.0050 \\
    Swin block4 (20k) & 0.8713 $\pm$ 0.0027 & 0.7914 $\pm$ 0.0029 & 0.4528 $\pm$ 0.0046 \\
    Swin block4 (200k) & 0.8850 $\pm$ 0.0027 & 0.8096 $\pm$ 0.0031 & 0.4272 $\pm$ 0.0048 \\\hline
    
    Custom Swin (20k) & 0.8640 $\pm$ 0.0022 & 0.7858 $\pm$ 0.0026 & 0.4684 $\pm$ 0.0035 \\
    Custom Swin (200k) & 0.8837 $\pm$ 0.0028 & 0.8077 $\pm$ 0.0030 & 0.4352 $\pm$ 0.0056 \\
    Custom Swin + MoCo (20k) & 0.8730 $\pm$ 0.0025 & 0.7950 $\pm$ 0.0035 & 0.4493 $\pm$ 0.0044 \\
    Custom Swin + MoCo (200k) & 0.8886 $\pm$ 0.0025 & 0.8126 $\pm$ 0.0023 & 0.4234 $\pm$ 0.0045 \\
    Custom Swin + MoCo (1.6M) & 0.8966 $\pm$ 0.0025 & 0.8222 $\pm$ 0.0026 & 0.4077 $\pm$ 0.0049 \\
    \bottomrule
    \end{tabular}
}
\end{table}

\subsection{Score--Multiplicity Correlation for All Models}
\label{sec:score_multiplicity_all}

\begin{figure}[hbt]
  \centering
  \includegraphics[width=7.7cm]{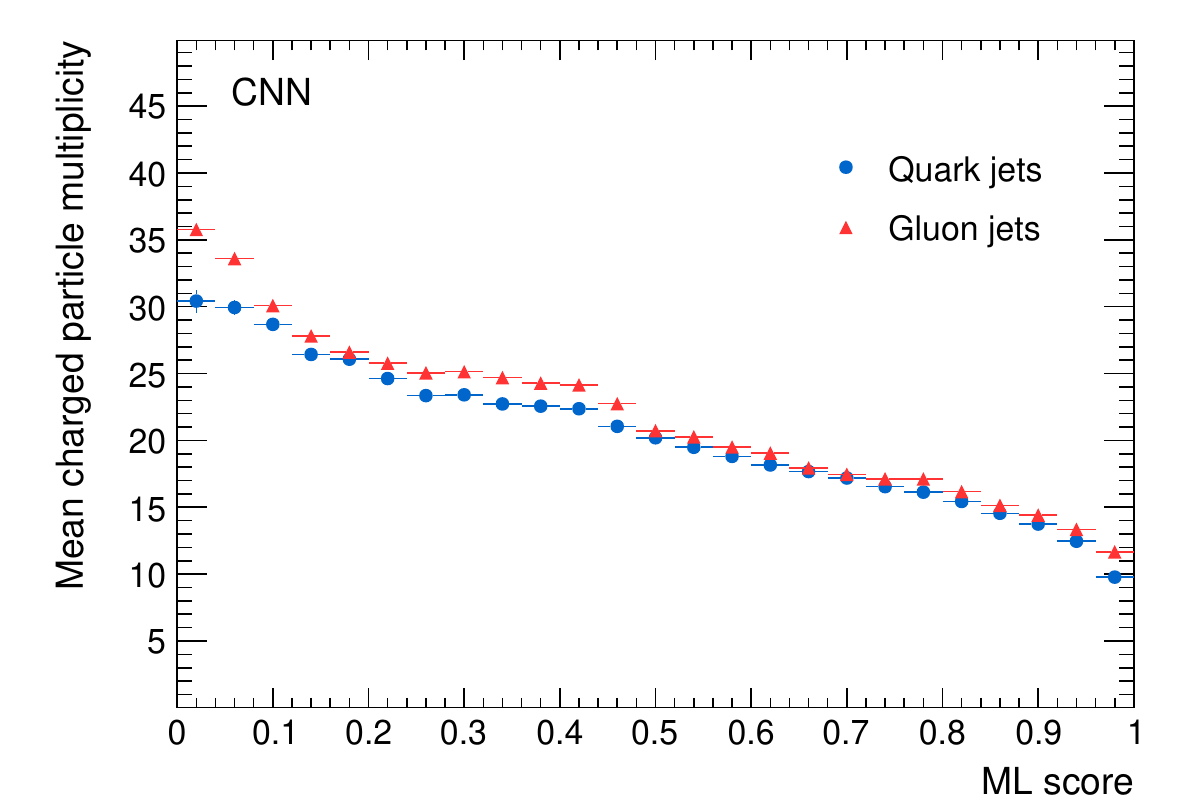}
  \includegraphics[width=7.7cm]{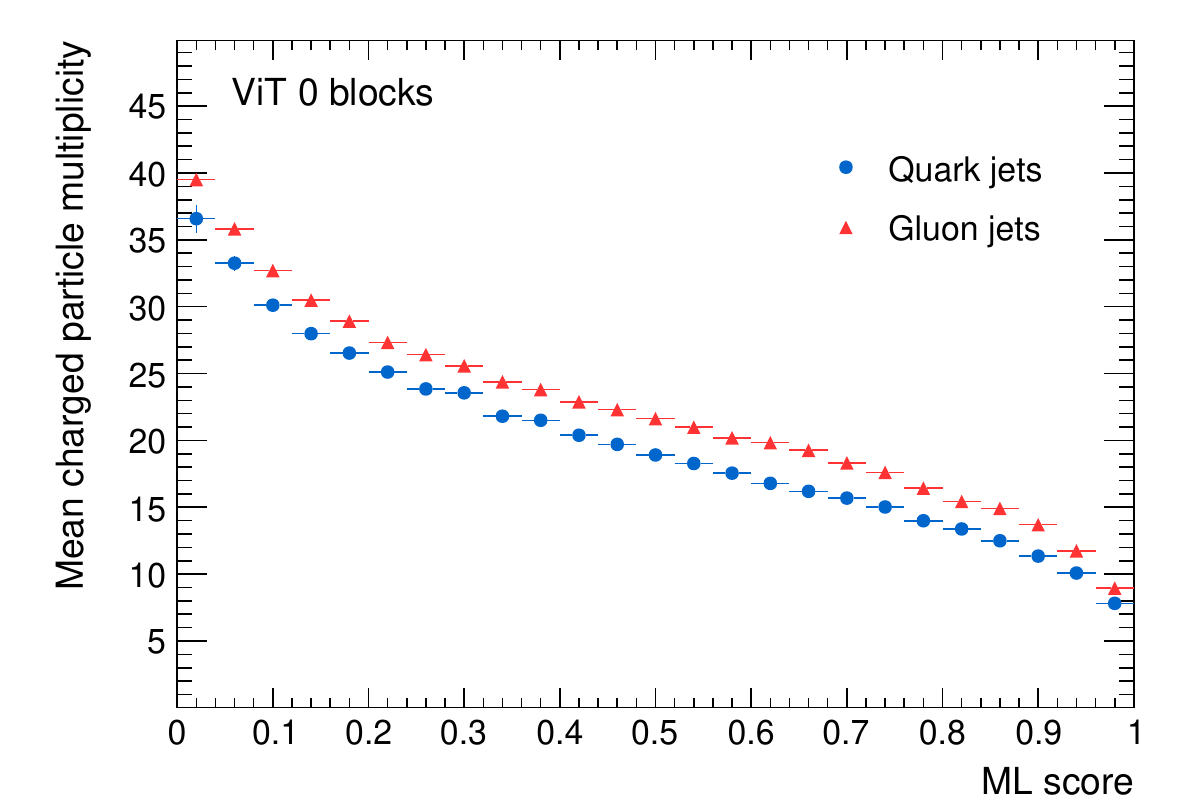}
  \includegraphics[width=7.7cm]{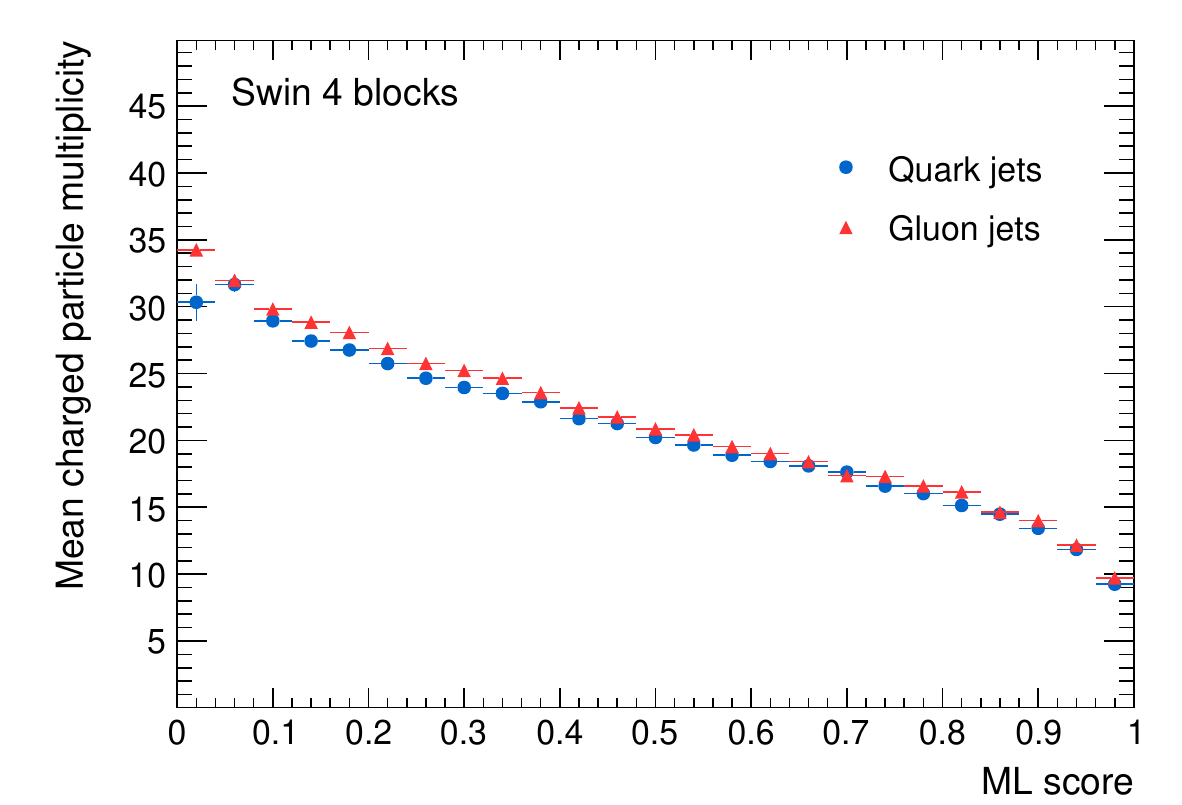}
  \includegraphics[width=7.7cm]{Figs/Figure15d_ScoreVsMult_CustomSwinMoCo_Rebinned.pdf}
  \caption{Mean charged-particle multiplicity as a function of the model score for quark and gluon jets obtained with CNN (top left), ViT with 0 unfrozen blocks (top right), Swin with 4 unfrozen blocks (bottom left) and custom Swin with MoCo (bottom right). The bottom-right panel shows the same model as Figure~\ref{fig:score_multiplicity} in the main text, repeated here for direct comparison.}
\label{fig:score_multiplicity_all}
\end{figure}

Figure~\ref{fig:score_multiplicity_all} shows the mean charged-particle multiplicity as a function of the model score for all four models. A similar trend is observed across the different architectures, with some differences in the detailed dependence on the model score.

\subsection{Model Architecture and Training}\label{app:modelHypParam}

\subsubsection{CNN Architecture and Training}
\label{app:cnn}
The CNN used in this study takes a $72\times72\times3$ jet image as input and consists of three convolutional blocks followed by a fully connected classifier. Each convolutional block is composed of a $3\times3$ convolution, batch normalization, ReLU activation, dropout, and $2\times2$ max-pooling. The number of output channels in the three convolutional blocks is 32, 64, and 128, respectively.
Since max-pooling is applied three times, the spatial resolution is reduced from $72\times72$ to $36\times36$, $18\times18$, and finally $9\times9$. Consequently, the final convolutional feature map has a dimension of $128\times9\times9$, which is flattened into a 10,368 dimensional feature vector.
The classifier consists of two fully connected layers. 
The first layer performs a dimensionality reduction from 10,368 to 128, followed by a ReLU activation and dropout. 
The final output layer maps 128 to 2 to produce logits for the two classes.
The total number of trainable parameters in this CNN architecture is 1,421,186.

The models were trained using the Adam optimizer with a learning rate of $10^{-3}$ and a batch size of 128 for 80 epochs. The best model was selected based on the validation AUC.

\subsubsection{ViT Architecture and Training}
\label{app:vit}
The ViT used in this study is based on the \texttt{vit\_tiny\_patch16\_224} model from the \texttt{timm} library, initialized with ImageNet-pretrained weights.
Since the jet images have a size of $72\times72\times3$, they were resized to $224\times224\times3$ before being fed into the model.
During fine-tuning, the patch embedding layer and positional embeddings were frozen, while only a subset of transformer blocks was updated. 
Three configurations were considered: 0 blocks (training only the classification head), 2 blocks (training the last two transformer blocks and the classification head), and 4 blocks (training the last four transformer blocks and the classification head).
The total number of parameters is 5,524,802, of which 770 (0 blocks), 890,498 (2 blocks), and 1,780,226 (4 blocks) parameters are trainable.

The models were trained using the Adam optimizer with a batch size of 128 for 80 epochs.
The learning rate was set to $10^{-4}$ for the 0 blocks configuration and $10^{-5}$ for the 2- and 4 blocks configurations. The best model was selected based on the validation AUC.
The detailed architecture configuration of the ViT is summarized in Table~\ref{tab:ViT_architecture}.

\begin{table}[t]
\centering
\caption{Architecture configuration of the ViT.}
\label{tab:ViT_architecture}
\begin{tabular}{lccccc}
\toprule
Stage & Spatial dimensions & Tokens & Dim & Blocks & Heads \\
\midrule
Patch Embedding & $14\times14$ & 196 & 192 & -- & -- \\
blocks 1--12     & $14\times14$ & 196(+CLS=197) & 192 & 12  & 3  \\
Head            & --           & --   & $192\rightarrow2$ & -- & -- \\
\bottomrule
\end{tabular}
\end{table}

\subsubsection{Swin Architecture and Training}
\label{app:swin}
The Shifted Window Transformer used in this study is based on the \texttt{swin\_tiny\_patch4\_window7\_224} model from the \texttt{timm} library, initialized with ImageNet-pretrained weights.
Since the jet images have a size of $72\times72\times3$, they were resized to $224\times224\times3$ before being fed into the model.
During fine-tuning, the patch embedding layer was frozen, while only a subset of Transformer blocks was updated. 
Three configurations were considered: 0 blocks (training only the classification head), 2 blocks (training the last two transformer blocks and the classification head), and 4 blocks (training the last four transformer blocks and the classification head).
The total number of parameters is 27,520,892, of which 3,074 (0 blocks), 14,186,930 (2 blocks), and 17,739,914 (4 blocks) parameters are trainable.

The models were trained using the Adam optimizer with a batch size of 128 for 80 epochs.
The learning rate was set to $10^{-4}$ for the 0 blocks configuration and $10^{-5}$ for the 2 and 4 blocks configurations. The best model was selected based on the validation AUC.
The detailed architecture configuration of the Swin Transformer is summarized in Table~\ref{tab:swin_architecture}.

\begin{table}[t]
\centering
\caption{Architecture configuration of the Swin Transformer.}
\label{tab:swin_architecture}
\begin{tabular}{lcccccc}
\toprule
Stage & Spatial dimensions & Tokens & Dim & Blocks & Heads & No. of windows\\
\midrule
Patch Embedding & $56\times56$ & 3136 & 96 & -- & -- & --\\
Stage 1         & $56\times56$ & 3136 & 96 & 2  & 3  & $8\times8=64$\\
Patch Merging   & $28\times28$ & 784  & 192& -- & -- & --\\
Stage 2         & $28\times28$ & 784  & 192& 2  & 6  & $4\times4=16$\\
Patch Merging   & $14\times14$ & 196  & 384& -- & -- & --\\
Stage 3         & $14\times14$ & 196  & 384& 6  & 12  & $2\times2=4$\\
Patch Merging   & $7\times7$ & 49   & 768& -- & -- & --\\
Stage 4         & $7\times7$ & 49   & 768& 2  & 24  & 1\\
Head            & --           & --   &$768\rightarrow2$& -- & -- & -- \\
\bottomrule
\end{tabular}
\end{table}

\subsubsection{Custom Swin Architecture and Training}
\label{app:custom_swin}

The custom Swin Transformer used in this study was designed to process jet images of size $72\times72\times3$ directly without resizing. 
A patch embedding layer with a patch size of $2\times2$ first converts the input image into $36\times36$ tokens with an embedding dimension of 64.
The backbone consists of three hierarchical stages with block numbers (2, 2, 4), embedding dimensions (64, 128, 256), and attention head numbers (2, 4, 8), respectively.
Within each stage, self-attention is computed inside local windows of size $3\times3$, and shifted-window attention alternates across consecutive blocks.

After the first and second stages, patch merging is applied to reduce the token resolution from $36\times36$ to $18\times18$, and then to $9\times9$, while increasing the embedding dimension from 64 to 128 and from 128 to 256. Each block consists of LayerNorm, window-based multi-head self-attention with residual connection, followed by another LayerNorm and a two-layer MLP with GELU activation. 
The hidden dimension of the MLP is set to four times the embedding dimension (MLP ratio = 4.0).
After the final stage, a LayerNorm layer is applied, followed by global average pooling over tokens and a linear classification head mapping 256 to 2. The detailed architecture configuration of the custom Swin Transformer is summarized in Table~\ref{tab:custom_swin_architecture}.

The model was trained from scratch using the Adam optimizer with a learning rate of $10^{-5}$, a batch size of 128, and 80 epochs. 
A dropout rate of 0.05 was used in the attention and MLP layers, and stochastic depth was applied with a maximum drop path rate of 0.2 that increases linearly with network depth. The total number of parameters is 3,823,886, and the best model was selected based on the validation AUC.

\begin{table}[t]
\centering
\caption{Architecture configuration of the custom Swin used in this study.}
\label{tab:custom_swin_architecture}
\begin{tabular}{lccccc}
\toprule
Stage & Spatial dimensions & Tokens & Dim & Blocks & Heads \\
\midrule
Patch Embedding & $36\times36$ & 1296 & 64 & -- & -- \\
Stage 1         & $36\times36$ & 1296 & 64 & 2  & 2  \\
Patch Merging   & $18\times18$ & 324  & 128 & -- & -- \\
Stage 2         & $18\times18$ & 324  & 128 & 2  & 4  \\
Patch Merging   & $9\times9$   & 81   & 256 & -- & -- \\
Stage 3         & $9\times9$   & 81   & 256 & 4  & 8  \\
Head            & --           & --   & $256\rightarrow2$ & -- & -- \\
\bottomrule
\end{tabular}
\end{table}

\subsubsection{MoCo Architecture and Training}
\label{app:moco}
MoCo pretraining was performed using the custom Swin backbone. 
The model consists of a query encoder and a momentum encoder, both sharing the same architecture. 
The backbone outputs a 256-dimensional feature, which is further mapped to a 128-dimensional contrastive representation using a two-layer projection head (256 $\rightarrow$ 256 $\rightarrow$ 128). 
Negative samples are drawn from a queue, with a queue size of 16,384. 
The momentum coefficient is set to 0.99, and the temperature parameter is set to 0.07.

The pretraining was conducted on the training split of the 200k jet image dataset.
The input images have a size of $72\times72\times3$, and two augmented views are generated from each image. No cropping is applied during augmentation. 
Random rotation by $0^\circ$, $90^\circ$, $180^\circ$, or $270^\circ$ is applied with a probability of 0.8.
Horizontal and vertical flips are applied with a probability of 0.5 each, and Gaussian blur is applied with a probability of 0.2, where the standard deviation of the Gaussian blur is uniformly sampled from $[0.1, 0.8]$.

The model is trained using the Adam optimizer with a learning rate of $10^{-4}$ and a batch size of 256.
Only the query encoder and the projection head are updated via backpropagation, while the momentum encoder is updated using a momentum update.
Pretraining and evaluation are conducted on different datasets. 
The query encoder is saved every 3 epochs, and a quick linear evaluation is performed at each checkpoint using a frozen backbone.
A linear classifier (256 $\rightarrow$ 2) is trained for 10 epochs, and the best model is selected based on the validation ROC-AUC, with early stopping applied.

\bibliographystyle{apsrev4-2}
\bibliography{reference}

\end{document}